\documentclass[12pt]{report}
\usepackage{lipsum} 
\usepackage[margin=1in]{geometry}
\usepackage{titlesec} 
\usepackage{amsmath}
\usepackage[table,xcdraw]{xcolor}
\usepackage{pdflscape}
\usepackage[export]{adjustbox}
\usepackage{array}
\usepackage{supertabular}
\usepackage{subcaption}
\usepackage{comment}
\usepackage{color}
\usepackage{graphicx}
\usepackage{multirow}
\usepackage{lscape}
\usepackage{diagbox}
\usepackage[colorlinks,citecolor=blue,urlcolor=blue,linkcolor=blue]{hyperref}
\usepackage[title]{appendix}
\usepackage{tikz}
\usepackage{multirow}
\usepackage{fancyhdr}

\titleformat{\chapter}[display]
  {\normalfont\Large\bfseries\centering}{\chaptertitlename\ \thechapter}{25pt}{\LARGE}

\definecolor{lime}{HTML}{A6CE39}

\newcommand{\empcirc}[2][black,fill=white]{\tikz[baseline=-0.5ex]\draw[#1,radius=#2] (0,0) circle ;} 
\newcommand{\fillcirc}[2][black,fill=black]{\tikz[baseline=-0.5ex]\draw[#1,radius=#2] (0,0) circle ;} 

\fancypagestyle{plain}{
  \fancyhf{} 
  \rhead{} 
  \lhead{Springer Studies in Computational Intelligence} 
  \lfoot{} 
}

\begin{document}

\setcounter{chapter}{9}
\chapter{Towards Smart Healthcare: Challenges and Opportunities in IoT and ML}
\vspace{-20pt}
\begin{center}
    Munshi Saifuzzaman \href{https://orcid.org/0000-0002-9236-5554}{\includegraphics[width=.125in,height=.125in,clip,keepaspectratio]{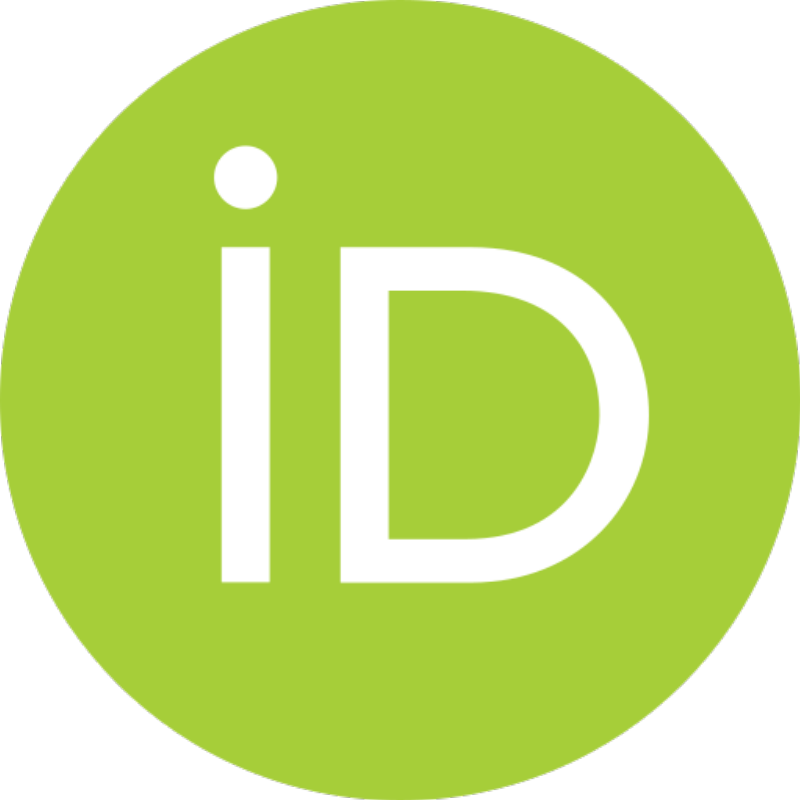}}\textsuperscript{1}, Tajkia Nuri Ananna \href{https://orcid.org/0000-0001-7385-980X}{\includegraphics[width=.125in,height=.125in,clip,keepaspectratio]{images/orcid.pdf}}\textsuperscript{2} \\ 
    \vspace{10pt}
    {\fontsize{14}{16} \textsuperscript{1} Dynamic Solution innovators, Dhaka 1206, Bangladesh\\
    \textsuperscript{2} Department of CSE, Metropolitan University, Sylhet, Bangladesh \\
     
    Email Address of the Corresponding Author: munshisaifuzzaman@gmail.com}
\end{center}

\section*{Abstract}

The COVID-19 pandemic and other ongoing health crises have underscored the need for prompt healthcare services worldwide. The traditional healthcare system, centered around hospitals and clinics, has proven inadequate in the face of such challenges. Intelligent wearable devices, a key part of modern healthcare, leverage Internet of Things technology to collect extensive data related to the environment as well as psychological, behavioral, and physical health. However, managing the substantial data generated by these wearables and other IoT devices in healthcare poses a significant challenge, potentially impeding decision-making processes. Recent interest has grown in applying data analytics for extracting information, gaining insights, and making predictions. Additionally, machine learning, known for addressing various big data and networking challenges, has seen increased implementation to enhance IoT systems in healthcare. This chapter focuses exclusively on exploring the hurdles encountered when integrating ML methods into the IoT healthcare sector. It offers a comprehensive summary of current research challenges and potential opportunities, categorized into three scenarios: IoT-based, ML-based, and the implementation of machine learning methodologies in the IoT-based healthcare industry. This compilation will assist future researchers, healthcare professionals, and government agencies by offering valuable insights into recent smart healthcare advancements.


\noindent\textbf{Keywords:} Healthcare analytics, machine learning integration, big data analytics, machine learning in healthcare, IoT systems in healthcare, research challenges, opportunities in healthcare.

\section{Introduction}
Smart healthcare refers to the implementation of cutting-edge technologies within the healthcare sector, including but not limited to the IoT, artificial intelligence (AI), machine learning (ML), deep learning (DL), and data analytics. Population expansion and a rise in the prevalence of diseases have combined to necessitate the development of a sophisticated healthcare system. In recent times, there have been numerous developments in the field of intelligent healthcare that have been observed globally. These include agile treatment, timely provision of services, remote monitoring of services, and timely response to emergency situations. Nonetheless, this presents the greatest obstacle to satisfying the rising demand for equipment and smart devices \cite{cook2012casas}. The resolution of this obstacle was achieved through the implementation of IoT in the healthcare sector. IoT implementation in the medical field has grown in popularity following advances in technology such as smart cities, smart regions, and smart devices.

The healthcare technology community has shown significant interest and anticipation in the IoT in recent years. With its practical applications, the healthcare industry stands to benefit from a wide range of opportunities due to IoT integration. The utilization of millions of sensors attached to a patient's body allows for the real-time collection of health data, enabling continuous remote health monitoring. Wireless Body Sensor Networks (WBSN) represent a key technology for patient monitoring \cite{alkhayyat2019wbsn}. These sensors gather crucial health data such as glucose levels, blood pressure, temperature, heart rate, and ECG readings \cite{abdullah2015real}. More recent systems incorporate actuators that can modify the external environment and enable notification or alarm systems. The application domain of IoT-based smart healthcare has seen remarkable advancements. This includes areas such as elderly monitoring, disease prediction, fitness tracking, remote monitoring, and disease treatment. These are just a few examples of the extensive developments in this field \cite{wu2019wearable,fu2015system, heshmat2018framework}.

Amid the advancements, a substantial amount of data, commonly referred to as big data, is generated by IoT devices. This data requires preprocessing to extract valuable insights for subsequent analysis. ML plays a crucial role in this process, particularly in IoT-based smart healthcare, where it's integrated to manage big data analytics \cite{mahdavinejad2018machine, qian2020orchestrating}. ML techniques handle extensive data, learning from it, training systems, facilitating enhanced decision-making, and refining treatment design. These techniques derive meaningful insights, uncover patterns, and reveal concealed information within large datasets. The key components of this layer encompass clustering, classification, association analysis, time series analysis, and outlier analysis \cite{mahdavinejad2018machine, babu2018survey}. The integration of ML with IoT devices simplifies the monitoring, management, and analysis of medical reports. ML algorithms excel at processing extensive biological data and quickly detecting specific patterns and mutations associated with different diseases. This capability can expedite the discovery of new therapeutic solutions.


\textbf{Chapter Motivation:} Researchers have extensively investigated smart healthcare components, methodologies, and technologies over the years, leading to a wealth of comprehensive discussions on the subject. These studies have proven beneficial to researchers in shaping their future contributions. Consequently, the pivotal question arises: what distinguishes this chapter from others, and why should readers dedicate their time to exploring the foundational concepts presented herein? To address this inquiry, a comparative analysis is outlined below:

\begin{itemize}
    \item To the best of our knowledge, there has been a notable absence of dedicated studies, including book chapters, focusing on smart healthcare challenges and future directions in recent years.
    \item Our investigation involved searches on prominent digital libraries utilizing the \textit{("survey" OR "review") on ("smart healthcare" AND ("Challenges" OR "Future Directions")) using (("IoT" OR "Internet of Things") AND ("ML" OR "Machine Learning"))} keywords. Among these, the most relevant review or survey studies are identified and summarized in Table \ref{tab:motivation}. The primary emphasis of our study is to discuss the existing and ongoing challenges faced by smart healthcare solutions and propose potential resolutions. The table illustrates that our chapter addresses the primary objectives that were lacking in previous studies. Notably, smart healthcare technologies that do not align with the primary motivation were deliberately excluded from consideration.
\end{itemize}
\begin{table}[h]
\centering
\caption{Comparison between recent studies and this work (applying of ML into IoT is represented as ML and IOT. \fillcirc{3pt} means yes and \empcirc{3pt} means no)}
\label{tab:motivation}
\resizebox{\columnwidth}{!}{%
\begin{tabular}{ll|cccccc}
\hline
\rowcolor[gray]{.9} 
\multicolumn{2}{l|}{\cellcolor[gray]{.9}\textbf{\diagbox{Comparison Type}{Recent Studies}}} & \multicolumn{1}{l}{\cellcolor[gray]{.9}\textbf{\cite{ghazal2021iot}}} & \multicolumn{1}{l}{\cellcolor[gray]{.9}\textbf{\cite{mohammadi2022applications}}} & \multicolumn{1}{l}{\cellcolor[gray]{.9}\textbf{\cite{chawla2020ai}}} & \multicolumn{1}{l}{\cellcolor[gray]{.9}\textbf{\cite{alshehri2020comprehensive}}} & \multicolumn{1}{l}{\cellcolor[gray]{.9}\textbf{\cite{tunc2021survey}}} & \multicolumn{1}{l}{\cellcolor[gray]{.9}\textbf{Our work}} \\ \hline \hline

\multicolumn{2}{l|}{Book chapter}  & \empcirc{3pt} & \empcirc{3pt} & \empcirc{3pt} & \empcirc{3pt} & \empcirc{3pt} & \fillcirc{3pt} \\ 
\rowcolor[gray]{.96} 

\multicolumn{2}{l|}{\cellcolor[gray]{.96}Technologies}  & \empcirc{3pt} & \empcirc{3pt} & \fillcirc{3pt} & \empcirc{3pt} & \empcirc{3pt} & \empcirc{3pt} \\ \hline

\multicolumn{1}{l|}{} & IoT based & \fillcirc{3pt} & \fillcirc{3pt} & \empcirc{3pt} & \empcirc{3pt} & \fillcirc{3pt} & \fillcirc{3pt} \\ 
\multicolumn{1}{l|}{} & \cellcolor[gray]{.96}ML based & \cellcolor[gray]{.96}\fillcirc{3pt} & \cellcolor[gray]{.96}\empcirc{3pt} & \cellcolor[gray]{.96}\fillcirc{3pt} & \cellcolor[gray]{.96}\empcirc{3pt} & \cellcolor[gray]{.96}\empcirc{3pt} & \cellcolor[gray]{.96}\fillcirc{3pt} \\ 
\multicolumn{1}{l|}{\multirow{-3}{*}{Applications}} & ML and IoT & \fillcirc{3pt} & \empcirc{3pt} & \empcirc{3pt} & \fillcirc{3pt} & \empcirc{3pt} & \fillcirc{3pt} \\ \hline

\multicolumn{1}{l|}{} & \cellcolor[gray]{.96}IoT based & \cellcolor[gray]{.96}\empcirc{3pt} & \cellcolor[gray]{.96}\empcirc{3pt} & \cellcolor[gray]{.96}\empcirc{3pt} & \cellcolor[gray]{.96}\empcirc{3pt} & \cellcolor[gray]{.96}\fillcirc{3pt} & \cellcolor[gray]{.96}\fillcirc{3pt} \\ 
\multicolumn{1}{l|}{} & ML based & \empcirc{3pt} & \empcirc{3pt} & \empcirc{3pt} & \empcirc{3pt} & \empcirc{3pt} & \fillcirc{3pt} \\ 
\multicolumn{1}{l|}{\multirow{-3}{*}{Challenges}} & \cellcolor[gray]{.96}ML and IoT & \cellcolor[gray]{.96}\empcirc{3pt} & \cellcolor[gray]{.96}\empcirc{3pt} & \cellcolor[gray]{.96}\empcirc{3pt} & \cellcolor[gray]{.96}\empcirc{3pt} & \cellcolor[gray]{.96}\empcirc{3pt} & \cellcolor[gray]{.96}\fillcirc{3pt} \\ \hline

\multicolumn{1}{l|}{} & IoT based & \empcirc{3pt} & \empcirc{3pt} & \empcirc{3pt} & \empcirc{3pt} & \fillcirc{3pt} & \fillcirc{3pt} \\ 
\multicolumn{1}{l|}{} & \cellcolor[gray]{.96}ML based & \cellcolor[gray]{.96}\empcirc{3pt} & \cellcolor[gray]{.96}\empcirc{3pt} & \cellcolor[gray]{.96}\empcirc{3pt} & \cellcolor[gray]{.96}\empcirc{3pt} & \cellcolor[gray]{.96}\empcirc{3pt} & \cellcolor[gray]{.96}\fillcirc{3pt} \\ 

\multicolumn{1}{l|}{\multirow{-3}{*}{\begin{tabular}[c]{@{}l@{}}Future Work\\  Directions\end{tabular}}} & ML and IoT & \empcirc{3pt} & \fillcirc{3pt} & \empcirc{3pt} & \fillcirc{3pt} & \empcirc{3pt} & \fillcirc{3pt} \\ \hline
\end{tabular}%
}
\end{table}

\noindent Readers are encouraged to delve into this chapter for a comprehensive and current understanding of the evolving landscape of smart healthcare.

\textbf{Contributions:} This chapter extensively discusses the obstacles that impede the progress of smart healthcare using IoT and ML approaches. The aim is to assist future contributors in understanding and addressing these challenges. Specifically, the contributions of this chapter can be summarized as follows:
\begin{enumerate}   
\item This chapter extensively discusses the existing smart healthcare applications, research challenges, and future work directions in three distinct scenarios: IoT-based, ML-based, and employing ML in IoT-based healthcare. 
\item While exploring existing smart healthcare applications, we not only provided brief details on these applications but also delved into a separate discussion on how ML has transformed this sector. Additionally, we highlighted the synergies between ML and IoT, showcasing how this combination has played a crucial role in enhancing healthcare systems.
\item Existing schemes have been explored in order to understand the challenges faced by current researchers developing ML-driven IoT applications.

    
    
\end{enumerate}

\textbf{Chapter Organization:} The chapter commences with Section \ref{sec: ml_iot_applications}, which explores the real-life applications of the three individual domains. This section starts by briefly outlining the applications of IoT-based healthcare (\ref{subsec:iot}), followed by examples of how ML has transformed the healthcare domain (\ref{subsec:ml}), and concludes by examining the integration of ML in IoT-based smart healthcare solutions (\ref{subsec:iot+ml}). The subsequent section (\ref{sec:research_challenges}) provides an in-depth survey summarizing the challenges encountered in IoT and ML-based smart healthcare. This section is divided into three subsections, with each subsection individually representing the three domain-specific smart healthcare open challenges. Finally, the chapter concludes by outlining domain-specific potential future directions in Section (\ref{sec:future}), as based on the survey.


\section{Exploring ML and IoT Applications}
\label{sec: ml_iot_applications}
Smart healthcare is a technologically advanced field that revolutionizes conventional medical and healthcare systems. Its primary objectives are to enhance healthcare services, improve patient care, and optimize the entire healthcare ecosystem through the integration of cutting-edge technologies, data-driven insights, and interconnected devices. By doing so, it aims to make healthcare more efficient, adaptable, and customized. Smart healthcare relies on a variety of key components, including IoT devices, wearable and implantable medical devices, AI, electronic health records (EHR), big data analytics, and more. It goes beyond clinical transformations, encompassing the collection, secure storage, and efficient processing of diverse physiological data. This comprehensive approach facilitates the early detection of diseases and even preventive measures against various medical conditions. One notable feature of smart healthcare is its capacity to facilitate remote patient monitoring through internet-connected devices. This capability is especially valuable for individuals with disabilities and the elderly. Moreover, it has the potential to significantly reduce healthcare costs while simultaneously enhancing the quality of life for patients \cite{yin2018smart}. The integration of technology in the healthcare sector, known as smart healthcare, is a notable advancement. This approach utilizes technological capabilities to enhance the quality, accessibility, and patient-centricity of healthcare services. Consequently, it contributes to the enhancement of health outcomes and the optimization of healthcare delivery efficiency. 

The services provided by smart healthcare based on the target user can be utilized in several categories, including clinical/scientific research institutions (e.g., hospitals), regional health decision-making institutions, or individual/family users/personalized service. This section provides a brief overview of several IoT-based smart healthcare applications, followed by a discussion of ML-integrated smart healthcare. It concludes with a demonstration of applications showcasing the intersection of IoT and ML in various domains.

\subsection{Practical Uses of IoT in Smart Healthcare}
\label{subsec:iot}
 The application of IoT-based smart healthcare can be divided into several based on different needs \cite{tian2019smart}. Various characteristics of IoT possess the potential to revolutionize the healthcare domain to a great extent. This subsection provides a brief overview of some particularly important examples of IoT-based smart healthcare.
 
\vspace{0.5cm}
\noindent \textit{Remote Monitoring and Disease Detection}\\
  IoT provides continuous remote monitoring through wearable devices, which plays a critical role in early disease detection and prevention. Continuous remote monitoring through wearable devices plays a critical role in early disease detection and prevention. 

  Wu et al. \cite{wu2019wearable} have presented an ECG data monitoring system that involves the attachment of a bipotential device to the user’s t-shirt, with data transmitted to a smartphone via Bluetooth. This setup empowers healthcare professionals to detect unusual disease signs, enabling timely intervention and treatment for patients.
  Agustine et al. \cite{agustine2018heart} have introduced an integrated alert system that issues notifications when oxygen levels surpass predefined thresholds. This alert system can facilitate prompt medical interventions, potentially saving lives.

\vspace{0.5cm}
\noindent \textit{Smart Treatment and Smart Surgical Environment}\\
IoT-based methodologies play a crucial role in cancer treatment, as demonstrated by Heshmat et al.'s novel technique integrating various stages like chemotherapy and radiotherapy. This system securely stores laboratory test data on a cloud-based server, empowering physicians to monitor and regulate prescription dosages. Moreover, it facilitates remote consultations via a dedicated mobile application \cite{heshmat2018framework}.

In surgical training and medical operations, IoT has been instrumental in developing advanced solutions like surgical robotics and virtual reality-based training environments. An exemplary advancement is a surgical training framework utilizing virtual reality to simulate authentic training scenarios, enabling global interaction among surgeons for collaborative learning and expertise exchange \cite{cecil2018iomt}. Notable examples of robotic systems include the Da Vinci system (manufactured by Intuitive Surgical, Sunnyvale, CA, USA), the Sensei X robotic catheter system (Hansen Medical, Auris Health, Inc., Redwood City, CA, USA), and the Flex® robotic system (Medrobotics, Raynham, MA, USA).

\vspace{0.5cm}
\noindent \textit{Virtual Health Platform and Assistant}\\
Intelligent healthcare introduces the concept of a readily accessible mHealth platform for medical professionals, patients, and researchers, facilitating remote patient monitoring and offering telemedicine services. This platform supports collaborative efforts for disease research. Additionally, smart healthcare systems empower patients to self-monitor their physical condition, as exemplified by the Stress Detection and Alleviation system using wearable medical sensors for continuous stress level tracking and autonomous stress reduction support \cite{estrin2010open, akmandor2017keep}.
IoT-based healthcare systems leverage virtual assistants as intermediaries, aiding patients in translating everyday language into medical terms and autonomously providing pertinent information to physicians, optimizing patient management, and enhancing medical procedures for more efficient care delivery and time savings.

\vspace{0.5cm}

\noindent \textit{Smart Hospitals and Pharmaceutical Industry}\\
Smart hospitals employ IoT-driven intelligent healthcare, utilizing advanced technologies to improve patient care and offer customized services for medical staff, patients, and administrators. These technologies enable functions like patient monitoring, daily medical personnel management, and tracking of biological specimens and medical instruments within hospital environments. Furthermore, smart healthcare in the pharmaceutical sector optimizes operations such as inventory management, anti-counterfeiting measures, and drug production, benefiting patients with streamlined access to physical examination systems, online appointment scheduling, and enhanced doctor-patient interactions. Automated processes expedite the patient's medical journey, while IoT-based solutions contribute to revolutionary advancements, aiding in disease research and conducting more effective clinical trials. In particular, continuous real-time monitoring through smart wearable devices proves valuable in trials related to lung disease, offering timely and precise information \cite{geller2016smart}.


\subsection{Practical Uses of ML in Smart Healthcare}
\label{subsec:ml}
ML and DL have the potential to enhance the intelligence of the smart healthcare domain by leveraging their ability to uncover novel insights from data. By applying scientific and mathematical techniques, ML can reveal hidden patterns in data, facilitating more informed decision-making. This capability can be utilized to ensure effective patient monitoring, detect diseases in advance, and enhance overall efficiency in smart healthcare. An example application involves the integration of electronic health records to identify patterns in infectious diseases, enabling the early detection of potential outbreaks. ML algorithms can play a crucial role in analyzing EHR data to swiftly identify individuals at risk, providing healthcare providers with real-time and more accurate alerts. This proactive approach can significantly improve response times and aid in preventing the spread of diseases more effectively. In addition, effectively managing the vast amount of data generated in the healthcare domain can only be achieved through the efficient application of ML. There exist multiple domains within smart healthcare where integrating ML can enhance the intelligence of healthcare systems. This subsection provides a brief overview of some particularly notable examples.

\vspace{0.5cm}
\noindent\textit{Remote Monitoring and  Emergency care}\\
Remote monitoring stands out as a prime feature in smart healthcare, revolutionizing the healthcare system. The integration of ML into healthcare systems significantly enhances accuracy and efficiency. In \cite{hassan2019hybrid}, Hassan et al. have proposed a context-aware framework, entitled "Hybrid Real-time Remote Monitoring," designed for remote patient monitoring, where authors employ a Naïve Bayes classifier in conjunction with the Whale Optimization algorithm. The collaborative use of these techniques aims to improve classification accuracy and achieve faster processing, thereby enhancing the efficiency of real-time patient monitoring. Monitoring and analyzing the patient's glucose and blood pressure readings, supervised learning techniques, particularly support vector machines (SVM), predict the presence of hypertension or abnormal levels of diabetes.

\vspace{0.5cm}
\noindent\textit{Disease Prediction and Prevention}\\
Within the expansive range of tasks that MLcan perform, prediction stands out as one of its most powerful capabilities. Whether predicting and determining the progression of diseases in patients with chronic illnesses \cite{pham2017predicting} or detecting mosquito-borne diseases earlier \cite{vijayakumar2019fog}, ML has shown the potential to revolutionize traditional healthcare, transforming it into intelligent healthcare. This transformative potential of ML in predicting and preventing diseases not only enhances the ability to forecast disease progression but also enables proactive measures for early detection and intervention.


\vspace{0.5cm}
\noindent\textit{Precision Medicine}\\
Precision medicine, also known as personalized medicine, is an innovative approach that utilizes individuals' genetic, environmental, and lifestyle information to recommend tailored medical interventions for each person \cite{precisionmedicine}. The integration of ML with DL has had a profound impact on advancing this field. For example, in \cite{dong2015anticancer}, Dong et al. have formulated and assessed a SVM model to predict the sensitivity of anticancer drugs using genomic data. They have proved that by leveraging genomics, it's possible to predict a patient's response to cancer treatment. If implemented in clinical settings, this approach could potentially spare non-responders from unnecessary treatments, directing them toward the most effective treatment based on their individual genome.

    
\vspace{0.5cm}
\noindent\textit{Decision Support Systems}\\
Decision support systems are increasingly utilized in the healthcare industry, enhancing the ability of doctors and hospitals to deliver improved treatment. While doctors remain the primary decision-makers, these systems enable them to expand their knowledge through health data, research databases, and examinations. Decision support systems based on ML can expedite decision-making, offer treatment suggestions, and provide justifications that help patients' family members understand the entire procedure. This not only allows doctors to allocate more time to communicating with patients but also alleviates the pressure of acquiring extensive knowledge and facilitates seeking a second opinion on decisions. A potential reference is available in \cite{ghazal2021iot}, where the authors propose initial therapies or treatments. This is particularly valuable in emergency situations where time constraints necessitate swift decision-making. Furthermore, by integrating bottleneck attention modules to distinguish between abnormal and normal DFU cases, a convolutional neural network (CNN)-based system for diabetic foot ulcers (DFU) has been suggested in \cite{das2023aespnet}.

\begin{figure}[t]
    \centering
    \includegraphics[width=0.6\textwidth]{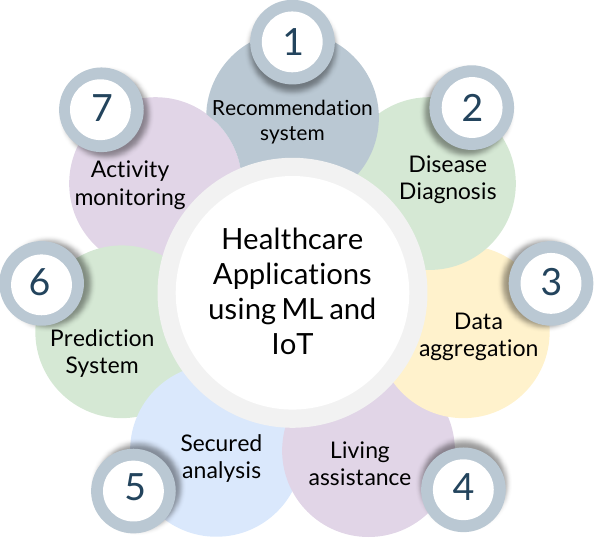}
    \caption{ML based IoT Applications}
    \label{fig:advantages}
\end{figure}

\subsection{Healthcare Applications Using ML and IoT}
\label{subsec:iot+ml}
Despite the significant advancements in IoT-based smart healthcare, several persistent challenges hinder its progress. One major obstacle is the deployment of IoT systems capable of managing massive amounts of data while ensuring robust security measures for data confidentiality, integrity, authorization, and authentication. The sensors and devices within these systems generate extensive data, often referred to as big data, characterized by highly correlated and redundant patterns. To address this issue, ML techniques have emerged as pivotal tools. ML possesses the capability to effectively handle vast volumes of data and extract meaningful insights from it \cite{farahani2020healthcare}. Various ML and DL techniques, such as convolutional neural networks, autoencoders (AE), deep belief networks (DBNs), long short-term memory (LSTM), and recurrent neural networks (RNNs), are being integrated to manage the enormous data generated in smart healthcare systems \cite{tuli2020healthfog, ahmad2020review}. This fusion of ML and IoT holds significant promise for immense progress in smart healthcare. Overall, the integration of ML techniques with IoT technologies is leading to significant improvements in the field of smart healthcare by addressing the challenges associated with vast data management and security concerns. With services such as real-time disease identification, prediction, and diagnosis, ML has the potential to revolutionize the healthcare industry and has critical importance in remote diagnosis. Additionally, AI-powered assistive systems facilitate care for trauma patients and aid in their recovery. This section presents some examples of some of the applications of ML in IoT-based smart healthcare. Fig. \ref{fig:advantages} presents a detailed visualization of the most prominent areas that can be benefited by the integration of ML and IoT in smart healthcare.

\vspace{0.5cm}
\noindent\textit{Recommendation System}\\
    In the realm of recommendation systems, many inventions have been proposed utilizing the fusion of IoT and ML. In \cite{asthana2017recommendation}, authors have proposed a recommendation system that uses IoT wearable devices to collect information such as previous history, demographic information, and retrieval of archived data from the sensors attached to the patient and applies various ML techniques, such as decision trees, logistic regression, and LibSVM, to predict the occurrence of diseases. Based on the information, a customized recommendation system is developed from the output. 
    
\vspace{0.5cm}
\noindent\textit{Data Aggregation}\\
    ML techniques are integrated to carry out efficient data aggregation, as data aggregation is a major step in smart healthcare. In \cite{qiu2016efficient}, authors have proposed a self-organized algorithm that aggregates healthcare data that has been collected via sensors. The proposed scheme reduces the high-dimensional space into a low-dimensional space that lowers the amount of transmitted data in the network and enhances the network lifetime, which also enhances the quality of the aggregated data.
    
\vspace{0.5cm}
\noindent\textit{Disease Diagnosis}\\
    By combining IoT and ML, it has become feasible to conduct a more precise and current disease diagnosis. An enhanced DL-assisted convolutional neural network (EDCNN) has been integrated into the Internet of Medical Things (IoMT) platform in \cite{pan2020enhanced}, thereby facilitating the diagnosis of cardiovascular disease. Furthermore, significant progress has been achieved in the identification of lung cancer through the implementation of sophisticated ML algorithms and IoT systems \cite{pradhan2020medical}. Active patient activity recognition and monitoring is another application of ML. Negra et al. (2018) employed ML techniques such as SVMs, linear discriminant analysis (LDA), and random forest to classify the activity of a patient \cite{negra2018wban}.
    
\vspace{0.5cm}
\noindent\textit{Living Assistance}\\
    The successful integration of IoT and ML in the field of ambient assisted living (AAL) has significantly enhanced the quality of life for individuals facing various disabilities and requiring constant care.
    In \cite{rupasinghe2022towards}, the authors have proposed a system that utilizes wrist-worn devices to collect data, employing a supervised ML algorithm, the Decision Tree Classifier. This algorithm can recognize four different activities: walking, sitting, sleeping, and standing. This real-time activity recognition serves as a constant assistant for individuals with disabilities or elderly individuals. The adoption of supervised ML techniques is assessed to overcome the challenges associated with real-time activity recognition.  

\vspace{0.5cm}
\noindent\textit{Secured Analysis}\\
  As a result of the sensitive nature of healthcare information, it is critical to ensure its confidentiality and security. The authors in \cite{khan2018performance} have proposed a highly secure system for an IoT-based healthcare environment, which utilizes a range of ML techniques to ensure the secure classification of patient data.
    
\vspace{0.5cm}
\noindent\textit{Prediction System}\\
    By integrating its predictive capabilities with IoT-based smart healthcare, ML is capable of developing an intelligent disease prediction system. The authors of \cite{elsaadany2017wireless} have proposed an IoT and ML-integrated cardiac arrest prediction system. The sensor acquires ECG signals containing information regarding cardiac activity. Following noise reduction, the acquired cardiac information is compared to a predetermined threshold in order to make a prediction regarding the outcome.
    Satija et al. \cite{satija2017real} have developed an algorithm for evaluating signal quality that is energy efficient. The algorithm analyzes ECG data and makes predictions about cardiovascular diseases (CVD) using ML techniques. 
    
\vspace{0.5cm}
\noindent\textit{Activity Monitoring}\\
    As mentioned earlier, IoT-enabled home healthcare systems represent a prevalent instance of smart healthcare solutions. The incorporation of ML technologies in this domain has demonstrated notable success. Authors from \cite{jan2019smartedge} introduced a system capable of detecting human presence without relying on cameras or motion sensors. This system gathers interaction data by monitoring activities like reading or writing across a diverse range of devices, subsequently leveraging multiple ML classification algorithms, including the C4.5 decision tree, linear support vector classifier (SVC), and random forest, to identify the presence of a human.


\section{Research Challenges}
\label{sec:research_challenges}
\begin{figure}
    \centering
    \includegraphics[width=0.75\textwidth]{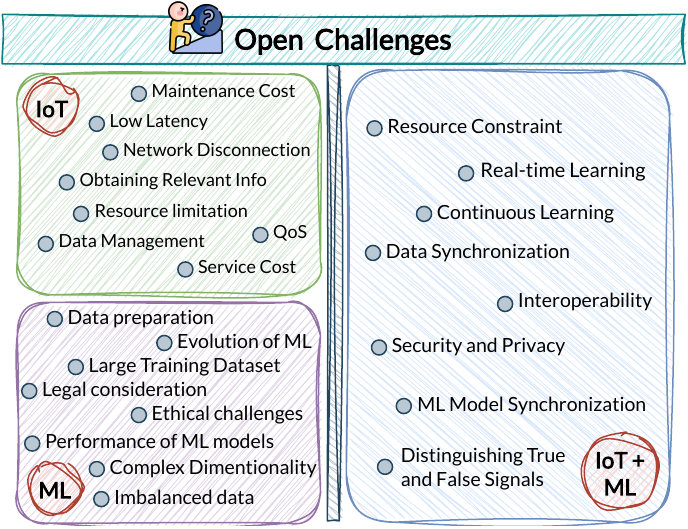}
    \caption{The Open Challenges building smart healthcare}
    \label{fig:challenges}
\end{figure}

In addressing the research challenges, the section is partitioned into two subsections, namely \textit{domain-specific challenges} and \textit{open research challenges}. Exploring domain-specific challenges involves an examination of recent studies in various healthcare fields, detailing the obstacles addressed within their respective domains. Regarding open research challenges, the primary issues in incorporating ML techniques into IoT-based smart healthcare are delineated. The challenges are vividly illustrated in Figure \ref{fig:challenges}.

\subsection{IoT based Systems}
While the integration of IoT brings a transformative shift to traditional healthcare systems, it also introduces a host of challenges that must be promptly addressed to fully harness the technology's potential. This section explores the array of challenges encountered by IoT-based smart healthcare solutions.

\vspace{0.5cm}
\noindent\textit{Maintenance and Service Cost}\\
A significant hurdle in IoT-based systems is the substantial cost associated with maintaining IoT devices. The implementation of IoT devices in smart healthcare, including sensors and wearable devices, often involves the use of expensive communication technologies and hardware tools. This results in elevated service and maintenance costs for developing these devices. However, the integration of IoT in healthcare aims to enhance medical care while reducing overall costs. Consequently, a crucial challenge emerges in overcoming this contradiction by creating devices and sensors that require minimal maintenance.

\vspace{0.5cm}
\noindent \textit{Network Disconnection and Low Latency}\\
Network disconnection and low latency tolerance represent significant challenges in any IoT-based system. These issues become even more critical in the context of smart healthcare, where the system is directly linked to a person's life. The use of diverse devices and the large volume of generated data can create obstacles to receiving timely information. This delay can lead to critical issues, especially in emergencies such as sudden changes in blood pressure or heart rate during remote monitoring. If data isn't transmitted promptly, patients may miss out on timely treatment, posing significant health risks \cite{tunc2021survey}. Additionally, for smooth operation, IoT devices require consistent network connectivity. Any disruptions in the network flow can impede the delivery of services, leading to similar issues as mentioned earlier. Given the numerous devices, data, and potential internet issues, ensuring these prerequisites in every case is challenging. Therefore, it is crucial to integrate backup options to address such challenges \cite{tissaoui2020uncertainty}.

\vspace{0.5cm}
\noindent\textit{Data Management}\\
Managing, analyzing, and processing data in the realm of smart healthcare poses significant challenges due to the diverse nature of IoT devices and the substantial volume of data they generate. In the field of smart healthcare, there's a constant influx of massive data every second. The challenges lie in efficiently collecting this data, establishing standardized formats and structures, utilizing suitable data models, and providing semantic descriptions of their content. Additionally, the prevalent approach of employing cloud-based solutions for handling and analyzing this data is raising concerns. These architectures are struggling to meet the computational demands and precise timing constraints associated with the diverse and extensive data generated in smart healthcare \cite{tissaoui2020uncertainty}.

\vspace{0.5cm}
\noindent \textit{Limitations in Energy, Computational, and Storage Resources}\\
In the domain of IoT-based smart healthcare, the integration of small wearable devices and sensors introduces a critical concern: power usage. Ensuring a continuous power source is essential for the uninterrupted operation of these devices. Striking a balance between the compact size of these devices and their energy consumption requirements presents a significant challenge. Meeting energy requirements becomes intricate as there is limited room to increase the device size to accommodate larger energy sources. Furthermore, restricted storage and computational capabilities pose obstacles to implementing complex operations, such as cryptographic models for security. Consequently, a persistent question remains: how to enhance the battery life of IoT devices while preserving their compact size? \cite{tissaoui2020uncertainty}

\vspace{0.5cm}
\noindent \textit{Obtaining Relevant Information}\\
A significant hurdle in IoT-based smart healthcare lies in obtaining accurate signals and data from the user's or patient's body. Wearable devices or sensors responsible for gathering diverse body measurements, such as respiratory rate, electrocardiogram (ECG) data, and blood pressure, need to be positioned correctly to efficiently capture the relevant information. For instance, identifying breathing abnormalities or monitoring respiratory rates poses a challenge due to the multitude of sounds generated by the upper body. Any misplacement of the sensor or device can result in the collection of inaccurate information. Consequently, distinguishing between various types of signals and selecting the appropriate one proves to be a challenging task \cite{malasinghe2019remote}.

\vspace{0.5cm}
\noindent\textit{Quality of Service}\\
The quality of service provided by an IoT-based healthcare system in fulfilling its tasks determines its primary performance metric. Numerous challenges arise in meeting the quality requirements of IoT-based applications, including energy efficiency, sensing data quality, network resource consumption, and latency. The quality of wearable devices or sensors used in this domain plays a crucial role in determining the system's accuracy and the relevance of the collected information \cite{tissaoui2020uncertainty}. Consequently, the most critical factors in designing a smart healthcare system are low latency, high response time, high scalability, and the integration of backup options.

\subsection{ML based Systems}
\label{sec:ml_challenge}
The incorporation of ML has had a profound impact on revolutionizing the healthcare domain, transforming it into a smart and intelligent system. However, implementing ML in the smart healthcare domain presents several challenges. This subsection provides a concise overview of the challenges associated with integrating ML into the realm of smart healthcare.

\vspace{0.5cm}
\noindent\textit{Preparation of Data for ML Algorithms}\\
    Preparing data for ML algorithms is a challenging task, particularly when dealing with physiological or health-related data. Such data is often collected from diverse sources and exists in unstructured or semi-structured formats. Therefore, prior to applying any ML algorithm, it is crucial to preprocess the data to avoid potential impacts on model training and classification accuracy. This involves the critical integration of data from different sources, addressing outliers, noise, missing and inconsistent data, and transforming it into a standardized format. Dimensionality reduction may also be necessary. Moreover, data may require preprocessing for storage efficiency or to uphold the quality of data mining processes \cite{hu2015efficient}. Given the nature and volume of healthcare data, this process is notably challenging. Additionally, incorporating patient-specific factors, which vary for each patient, poses a considerable challenge to achieve.

\vspace{0.5cm}
\noindent \textit{Imbalanced Data}\\
Class imbalance is a prevalent factor in ML, occurring when the distribution of classes within a dataset is uneven. Specifically, one class comprises a significantly larger number of records than the other. In such cases, the implementation of any ML model tends to be biased toward the majority class. In healthcare, because ML assumes that classes are normally distributed, this bias could mean that the model does better at predicting outcomes for more common health conditions but worse on the minority class, which could potentially include critical conditions \cite{bellazzi2008predictive}. Furthermore, if the model exhibits bias toward a specific class, it can impact decision-making, lead to false alarms, overlook rare conditions, and ultimately have implications for patient health.

\vspace{0.5cm}
\noindent \textit{Complex Dimentionality and Large Training Dataset}\\
Data collected in the healthcare sector is typically high-dimensional, and while high dimensionality often provides relevant information for understanding patient conditions, applying ML algorithms to such data may lead to lower accuracy. Additionally, dealing with the vast amount of healthcare data poses a significant challenge. ML algorithms require an adequate amount of accurate data to effectively fulfill their roles. For instance, the application of DL to image-based health data demands a substantial and high-quality training dataset. Ensuring the validity and accuracy of the images is crucial. However, obtaining such image data, meeting both quantity and quality requirements, remains a challenging task \cite{chen2019develop}. Furthermore, any increase in the size of the training data contributes to the memory complexity of the model.

\vspace{0.5cm}
\noindent\textit{Performance of ML models}\\
Creating an effective model that performs well across diverse healthcare datasets is a worldwide challenge. For instance, convolutional neural networks (CNNs) excel in image-related tasks, while recurrent neural networks are proficient in waveform analysis. However, in the context of smart healthcare, where data is sourced from various origins, the application of a specific model may not necessarily yield optimal performance. Several factors contribute to the overall performance of ML implementations in smart healthcare. These include:
\begin{enumerate}
    \item Dimensionality: The high dimensionality of data can impact the effectiveness of ML models, especially when dealing with complex and varied healthcare datasets. 
    \item Noisy Data: The presence of noise or irrelevant information in the data can hinder the accuracy and reliability of ML algorithms.
    \item Redundant Data: Duplicate or redundant data may lead to inefficiencies in model training and may not contribute substantially to the learning process.

\item Outliers: Outliers can significantly influence the performance of ML models by skewing the training process and affecting the generalization of the model.

\item Number of Attributes: The number of attributes or features in datasets can pose challenges, especially when dealing with a large and varied set of healthcare data.

\end{enumerate}

\vspace{0.5cm}
\noindent\textit{Addressing Ethical and Legal Challenges} \\
Entrusting a ML model with the entire responsibility of processing raw medical data is challenging. Involving a medical professional becomes necessary to categorize and interpret the medical data. Numerous ethical and legal concerns surround the implementation of ML in healthcare. For instance, the output of a DL algorithm applied to healthcare data can be challenging to explain logically. Lack of involvement from a medical professional in such scenarios may pose severe threats and potential harm.

\vspace{0.5cm}
\noindent\textit{Evolution of ML with changing infrastructure}\\
Healthcare facilities adapt to patient demands, evolving in management, infrastructure, data, and training requirements. The dynamic nature of healthcare advancement poses a crucial question: Will the implemented ML model align and evolve with changing infrastructure while maintaining its original prediction logic? \cite{amador2022early} Addressing this challenge promptly is imperative.

\subsection{IoT and ML based Systems}
IoT and ML have collaborated to bring about significant advancements in the field of smart healthcare. However, there are still a lot of obstacles to overcome because of the inherent complexities of the individual fields as well as the complications that arise when combining these two fields. This section discusses the open research challenges that are associated with ML and IoT-based smart healthcare systems.

\begin{table}[h]
\centering
\caption{Open research challenges}
\label{tab:open_challenges}
\resizebox{\textwidth}{!}{%
\begin{tabular}{ll|ll|c}
\hline
\rowcolor[gray]{.96} 
\multicolumn{2}{l|}{\cellcolor[gray]{.96}\textbf{Challenges Type}} & \multicolumn{2}{l|}{\cellcolor[gray]{.96}\textbf{Issues}} & \multicolumn{1}{c}{\cellcolor[gray]{.96}\textbf{Relevant work}} \\ \hline \hline
\multicolumn{2}{l|}{} & \multicolumn{2}{l|}{Transmission of Correlated data} & \cite{naha2020deadline, zhou2017security} \\ 
\multicolumn{2}{l|}{} & \multicolumn{2}{l|}{\cellcolor[gray]{.96} Resource Management Agreement} & \cellcolor[gray]{.96} \cite{ali2020resource} \\ 
\multicolumn{2}{l|}{\multirow{-3}{*}{Resource Scarcity}} & \multicolumn{2}{l|}{Acceptable level of accuracy} & \cite{khan2020challenges} \\ \hline
\rowcolor[gray]{.96}  
\multicolumn{2}{l|}{\cellcolor[gray]{.96} } & \multicolumn{2}{l|}{\cellcolor[gray]{.96} Compromisation of transmitted data} & \cite{sharma2020performance, jan2019payload} \\ 
\multicolumn{2}{l|}{\cellcolor[gray]{.96} } & \multicolumn{2}{l|}{Management of heterogenous data} & \cite{flynn2020knock, williams2016always} \\ 
\rowcolor[gray]{.96}  
\multicolumn{2}{l|}{\multirow{-3}{*}{\cellcolor[gray]{.96} Security and Privacy}} & \multicolumn{2}{l|}{\cellcolor[gray]{.96} Existing techniques infeasibility} &  \\ \hline
\multicolumn{2}{l|}{Interoperability} & \multicolumn{2}{l|}{} & \cite{khan2014fairness, qadri2020future} \\ 
\rowcolor[gray]{.96}
\multicolumn{2}{l|}{Distinguishing True and False Signals} & \multicolumn{2}{l|}{} & \cite{malasinghe2019remote} \\ 
\multicolumn{2}{l|}{Data and ML Model Synchronization} & \multicolumn{2}{l|}{} & \cite{hassan2018intelligent} \\ \hline
\rowcolor[gray]{.96}  
\multicolumn{2}{l|}{\cellcolor[gray]{.96} } & \multicolumn{2}{l|}{\cellcolor[gray]{.96} Limited energy supplies} & \cite{park2020energy} \\ 
\multicolumn{2}{l|}{\cellcolor[gray]{.96} } & \multicolumn{2}{l|}{Success of the underlying applications} & \cite{abbasian2020survey} \\ \cline{3-5} 
\rowcolor[gray]{.96}  
\multicolumn{2}{l|}{\cellcolor[gray]{.96} } & \multicolumn{1}{l|}{\cellcolor[gray]{.96} } & Energy Algorithms & \cite{selvaraj2020challenges, mittal2019energy} \\ 
\multicolumn{2}{l|}{\multirow{-4}{*}{\cellcolor[gray]{.96} Energy management}} & \multicolumn{1}{l|}{\multirow{-2}{*}{\cellcolor[gray]{.96} Optimization}} & Routing Approach &  \\ \hline
\multicolumn{2}{l|}{} & \multicolumn{2}{l|}{Network performance} & \multicolumn{1}{l}{} \\ 
\multicolumn{2}{l|}{} & \multicolumn{2}{l|}{\cellcolor[gray]{.96}Ways of gaining insights} & \cellcolor[gray]{.96} \cite{gill2019bio} \\ 
\multicolumn{2}{l|}{} & \multicolumn{2}{l|}{Innovative noise removal techniques} & \cite{wan2019similarity} \\ \cline{3-5} 
\multicolumn{2}{l|}{} & \multicolumn{1}{l|}{\cellcolor[gray]{.96} } & \cellcolor[gray]{.96} Underlying topologies & \cellcolor[gray]{.96} \cite{qi2019convolutional, li2020secrecy, wiens2019engine, li2020uav} \\ 
\multicolumn{2}{l|}{\multirow{-5}{*}{Big data Analytics}} & \multicolumn{1}{l|}{\multirow{-2}{*}{\cellcolor[gray]{.96} Performance}} & \begin{tabular}[c]{@{}c@{}}Dynamics and \\diverse environments\end{tabular} & \cite{shokri2019review, xue2019using, ma2019numerical, khan2020secured, tingting2019three} \\ \hline
\end{tabular}%
}
\end{table}

\subsubsection{Open Research Challenges}
This section discusses the open research challenges in the domain of IoT and ML-based smart healthcare. The summarization is represented in Table \ref{tab:open_challenges}.

\vspace{0.5cm}
\noindent\textit{Resource Constraint}\\
The scarcity of resources, including sensors, devices, actuators, and microcontrollers, presents the greatest obstacle for the development of smart healthcare. Because of their relatively tiny size, these devices have limited processing capacity and poor computational capability. As a result, it is difficult to manage the resources by making sure that they are being used effectively reference \cite{khan2020challenges, ishtiaq2019performance, hussain2020machine}. In addition, the data that is produced by these devices with limited resources is strongly correlated with one another, redundant, and contains patterns that are quite similar to one another. Because of this, transmitting various kinds of data over the network for the purpose of analysis or storage consumes a significant amount of energy, which in turn lowers the quality of service and results in low throughput. This problem of scarce resources has been mitigated to some extent as a result of the integration of cloud services with the IoT; however, integration of cloud services leads to increased complexity, greater costs, and a higher level of required maintenance. In addition to this, these data come from a wide variety of sources, the quality of which may vary, as well as difficulties such as noisy data, inconsistent data, and a great number of other problems. The aforementioned challenges render ML-based data aggregation methods incapable of maintaining the integrity of the data, thereby resulting in higher energy consumption. The majority of current ML-based data aggregation methods are not energy efficient as a result of these concerns. 

In addition, effective management of resources is a critical challenge in smart healthcare. Because of the unique qualities of IoT networks, a number of issues related to resource management, including resource discovery, modeling, provisioning, scheduling, estimation, and monitoring, continue to be of greater importance than they were in the past \cite{ali2020resource}. In addition, the process of allocating resources is not nearly as efficiently optimized as it should be to increase the overall quality of the service.

\vspace{0.5cm}
\noindent\textit{Real-time and Continuous Learning}\\
In smart healthcare, sensors continuously gather real-time data, which is then fed into ML algorithms for analysis. This scenario persists as data is captured moment by moment in real time. Continuous learning, or online ML, involves learning from the ever-increasing data while retaining knowledge from previous datasets. While online ML appears to be an impactful solution, it has constraints, one of which is catastrophic forgetfulness. In this phenomenon, the ML model suddenly forgets what it has learned, leading to a significant reduction in performance \cite{hassabis2017neuroscience}.

\vspace{0.5cm}
\noindent\textit{Data and ML Model Synchronization}\\ 
In the landscape of IoT and ML-based smart healthcare, data from heterogeneous devices and sensors undergoes collection and transmission to the cloud for ML algorithm application and analysis. However, the heterogeneity in sensor internal clock structures introduces challenges in synchronization, necessitating the implementation of smart gateways. It becomes imperative to synchronize the collected data on a temporal basis before transmitting it to the cloud for preprocessing, model training, and development. In cases lacking edge devices, this entire process unfolds in the cloud, where decisions and alerts are generated. Any disruption in network and cloud services poses a severe threat to patients, potentially endangering their lives. To mitigate this risk, it is essential to design the system in a way that ensures data synchronization on local servers. For instance, Hassan et al. proposed a hybrid model, acknowledging a drawback involving the downloading and copying of the ML model from the cloud to local devices \cite{hassan2018intelligent}. Addressing this significant challenge is crucial for achieving optimal outcomes in smart healthcare systems.

\vspace{0.5cm}
\noindent\textit{Security and Privacy}\\
Integration of IoT into healthcare has enabled a number of benefits that were unimaginable just a few years ago, including individualized and immediate access to medical care, remote and continuous monitoring, and more. This has been achieved through the collaboration of healthcare devices and technology, which provide users with an extensive array of healthcare services. Between 2023 and 2028, the global healthcare IoT market is projected to expand by 12.32\%, culminating in a valuation of around $\$178$ billion by 2028\footnote{Healthcare IoT - Worldwide, available at: \href{https://www.statista.com/outlook/tmo/internet-of-things/healthcare-iot/worldwide}{www.statista.com}.}. Despite the ongoing and revolutionary progress in smart healthcare, the security and privacy implications of the implemented solution must be taken into account due to the sensitive nature of health data \cite{kaur2020internet, almolhis2020security, bansal2020iot}. The transmission of these data upstream not only has detrimental effects on the data aggregation techniques that are foundational to the system but also harms its overall performance. This leaves the network susceptible to a variety of threats, including denial of service, eavesdropping, Sybil, sinkhole, and sleep deprivation attacks. This continues to be a threat as the network continues to expand and a greater number of hardware and software vulnerabilities are introduced \cite{sharma2020performance}.

Furthermore, the inclusion of personally identifiable information in healthcare data, including but not limited to personal details, family history, electronic medical records, and genomic data, gives rise to concerns regarding the sensitivity of the data and its owner. An estimated 72\% of malicious traffic is directed at healthcare data with the intention of exploiting the system \cite{jan2019payload}. Hence, it is critical to ensure the confidentiality and protection of this data against hackers through the implementation of diverse security and privacy protocols \cite{bhattacharjya2020present}. In addition to these, additional obstacles may include inadequate physical security, misconfigured devices, or network vulnerabilities. Furthermore, the task of safeguarding data privacy and security is considerably complicated by the fact that the majority of the devices within the system are heterogeneous and are administered by third parties \cite{flynn2020knock, williams2016always}. Due to the limited resources inherent in IoT devices, the existing security features are not feasible enough to mitigate the issues.

\vspace{0.5cm}
\noindent\textit{Interoperability}\\
The combination of ML and the IoT has sped up advancements in medical care; however, the real obstacle is the absence of global standards that are recognized and approved of by all relevant authorities. Because of the diverse range of applications and application domains, it is becoming an increasingly difficult task to maintain a global collaborative environment, whether the topic at hand is the selection of technologies or algorithms \cite{khan2014fairness, qadri2020future}. Increased throughput, decreased unplanned outages, and lower maintenance costs are some of the advantages that come along with the use of devices that are interoperable. 



\vspace{0.5cm}
\noindent \textit{Distinguishing True and False Signals}\\
Various wearable devices play a crucial role in smart healthcare, employing different ML algorithms based on the application. However, the challenging aspect lies in accurately distinguishing between genuine and false signals. For instance, in a fall detection system, an accelerometer is typically utilized to identify falls, with a ML algorithm analyzing the data and sending alerts to caregivers. Yet, differentiating between a fall and a rapid movement or regular activities poses a formidable challenge. The system must exhibit robustness to discern any sudden movement, such as bending down or picking up an object, to prevent the introduction of false data into the ML algorithm, thereby avoiding the generation of inaccurate alerts. Furthermore, enhancing research data for fall detection is challenging due to the infrequency of such incidents \cite{malasinghe2019remote}.

\vspace{0.5cm}
\noindent\textit{Big Data Analytics}\\
Managing the enormous quantities of data that are consistently produced by connected medical devices is one of the most significant and crucial challenges that smart healthcare has to overcome. It is anticipated that the growth of the IoT in the future will be more exaggerated, which means that a greater quantity of data will be produced that is unstructured, raw, and highly correlated. These data are passed through the network and are used for analysis as well as decision-making. The data may contain correlations, redundant values, or null values. This has a negative impact on the performance of the network and makes it more vulnerable to a wide variety of different kinds of attacks \cite{yang2020federated}. In addition to this, gaining useful insights from these data is a challenging task for a variety of ML algorithms because it requires extensive preprocessing of data, which takes a lot of time, and managing this massive amount of data is very difficult \cite{gill2019bio}. It is essential to have this in place so that a variety of ML and DL techniques can be utilized to facilitate improved decision-making. In addition, because IoT devices produce real-time data, it is extremely difficult to apply ML algorithms to real-time data and carry out a variety of operations in order to respond appropriately.

At present, ML-based data aggregation techniques are inadequately optimized in terms of the nature of the data, which hinders their ability to identify outliers while maintaining service quality and efficiency. Moreover, the correlation between data aggregation and the fundamental topology of the network is more evident. The efficacy of these methodologies is significantly influenced by the foundational topologies \cite{qi2019convolutional,li2020secrecy,wiens2019engine}. To improve the quality of aggregated data and enhance data signal quality, it is critical to reduce noise, which is quite challenging to accomplish with current methods given the nature of the data. Furthermore, existing methodologies fail to adequately support the execution of diverse data analysis operations in a heterogeneous setting, a significant obstacle given the substantial influence of heterogeneity in the domain of intelligent healthcare \cite{shokri2019review, xue2019using, khan2020secured}.

\begin{table}[t]
\centering
\caption{Challenges considered in various healthcare domains (\fillcirc{3pt} means considered and \empcirc{3pt} means does not)}
\label{tab:challenges_paperwise}
\resizebox{\textwidth}{!}{%
\begin{tabular}{l|cccccc}
\hline
\rowcolor[gray]{.9} 
\textbf{\diagbox{Challenges \\ Considered}{(Metric)\\Reference}} & \textbf{\begin{tabular}[c]{@{}c@{}}(Sensor Level)\\\cite{asthana2017recommendation} \end{tabular}} & \textbf{\begin{tabular}[c]{@{}c@{}}(AR)\\ \cite{hsu2017fallcare+}\end{tabular}} & \textbf{\begin{tabular}[c]{@{}c@{}}(EEG \\information)\\ \cite{zhang2018internet} \end{tabular}} & \textbf{\begin{tabular}[c]{@{}c@{}}(CLA)\\ \cite{de2016wearable}\end{tabular}} & \textbf{\begin{tabular}[c]{@{}c@{}}(Diagnosis \\system)\\\cite{ara2017case}\end{tabular}} & \textbf{\begin{tabular}[c]{@{}c@{}}(H-IoT QoS)\\ \cite{fafoutis2018extending}\end{tabular}} \\ \hline \hline
Feature extraction & \fillcirc{3pt} & \empcirc{3pt} & \empcirc{3pt} & \empcirc{3pt} & \empcirc{3pt} & \fillcirc{3pt} \\
\rowcolor[gray]{.96} 
Cost-effectiveness & \fillcirc{3pt} & \empcirc{3pt} & \empcirc{3pt} & \empcirc{3pt} & \fillcirc{3pt} & \fillcirc{3pt} \\
Personalisation & \fillcirc{3pt} & \empcirc{3pt} & \empcirc{3pt} & \empcirc{3pt} & \empcirc{3pt} & \empcirc{3pt} \\
\rowcolor[gray]{.96} 
Efficiency & \fillcirc{3pt} & \empcirc{3pt} & \fillcirc{3pt} & \fillcirc{3pt} & \fillcirc{3pt} & \fillcirc{3pt} \\
Usefulness & \fillcirc{3pt} & \empcirc{3pt} & \empcirc{3pt} & \empcirc{3pt} & \empcirc{3pt} & \empcirc{3pt} \\
\rowcolor[gray]{.96} 
Big data analytics & \fillcirc{3pt} & \fillcirc{3pt} & \fillcirc{3pt} & \fillcirc{3pt} & \fillcirc{3pt} & \fillcirc{3pt} \\
Scalability & \empcirc{3pt} & \fillcirc{3pt} & \empcirc{3pt} & \empcirc{3pt} & \empcirc{3pt} & \empcirc{3pt} \\
\rowcolor[gray]{.96} 
Maintainability & \empcirc{3pt} & \fillcirc{3pt} & \empcirc{3pt} & \empcirc{3pt} & \empcirc{3pt} & \fillcirc{3pt} \\
\begin{tabular}[c]{@{}l@{}}Environmental factors\\ and sentiment status\end{tabular} & \empcirc{3pt} & \empcirc{3pt} & \fillcirc{3pt} & \empcirc{3pt} & \empcirc{3pt} & \empcirc{3pt} \\
\rowcolor[gray]{.96} 
Real-time data & \empcirc{3pt} & \empcirc{3pt} & \fillcirc{3pt} & \fillcirc{3pt} & \empcirc{3pt} & \empcirc{3pt} \\
Security and Privacy & \empcirc{3pt} & \empcirc{3pt} & \empcirc{3pt} & \empcirc{3pt} & \fillcirc{3pt} & \empcirc{3pt} \\ \hline
\end{tabular}%
}
\end{table}

\subsubsection{Challenges in Existing Systems}
In this section, various domains have been examined within the realm of H-IoT to ascertain the challenges encountered by researchers when proposing a new architecture. Specifically, recent literature has been reviewed encompassing different metrics, including sensor-level data, augmented reality, electroencephalogram (EEG) information, cognitive load assessment, diagnostic systems, and H-IoT quality of service (QoS). The identified challenges have been organized and presented in Table \ref{tab:challenges_paperwise}.


Asthana et al. \cite{asthana2017recommendation} have proposed a recommendation system that employs ML to classify input data and associate disorders with corresponding wearable devices. Their architectural design addresses the extraction of an individual's health conditions and the necessary measurements for monitoring. Furthermore, identifying the most cost-effective set of wearable devices for measuring these monitored parameters is a challenging endeavor, especially given the substantial proliferation of options in the contemporary market.


The field of IoT smart healthcare encompasses fall-detection services. FallCare faces three prominent issues: (1) constrained computational resources on the Raspberry Pi 2 (RPi 2); (2) a lack of live streaming capability for detection, relying on images, fall video clips, or event notifications; and (3) poor scalability and maintainability for updating learning models across distributed devices. In response to these challenges, the authors of \cite{hsu2017fallcare+} have introduced CTPhone, a fall identification system based on a smart home sensor network.

With widespread connectivity, brain-computer interfaces (BCI) have the potential to empower individuals to directly control objects, such as smart home appliances or assistive robots, using their thoughts. In the domain of EEG metrics, Zhang et al. \cite{zhang2018internet} have proposed a unified DL framework aimed at enabling human-thing cognitive interactivity. Specifically, the authors utilized EEG signals for speech generation and automation. Nevertheless, challenges pertaining to the accuracy of signal interpretation and the time-consuming nature of these tasks impede the realization of this vision.Raw brain signals can be acquired using various technologies, including electroencephalography, functional near-infrared spectroscopy (fNIR), and magnetoencephalography (MEG). These signals often exhibit low fidelity, are susceptible to noise and concentration issues, and therefore pose challenges for accurate signal interpretation. Furthermore, the preprocessing of brain signals and subsequent feature engineering are both time-consuming and heavily dependent on human domain expertise.

In \cite{de2016wearable}, Arruda et al. have proposed a model for measuring motion using micro-electro-mechanical systems (MEMS) technology, and they determined the device's location based on the measured signal. They have extracted the data by addressing time-delayed movements and random signal lengths before extracting features from the windowed signals. To enhance system accuracy and reduce computational complexity, Arruda et al. have implemented a feature selection algorithm to decrease the amount of data requiring processing. The improvement in accuracy was contingent on the choice of various classifiers, adjustments to the feature selection algorithm, or modifications to the window length.

Ara et al. \cite{ara2017case} have conducted a case study to assess the feasibility of integrating IoT and ML for enhancing the diabetes management system. Their vision was to design a cost-effective system capable of transmitting large volumes of generated data to the cloud. Furthermore, the authors have aimed to create a versatile tool that is easy to develop and proficient in data analysis and presentation.

In a related study by Fafoutis et al. \cite{fafoutis2018extending}, the authors have sought to extend sensor battery life. To achieve this, the model needed to be energy-efficient. Consequently, they have developed an SVM classifier that effectively segregates essential information from redundant data, resulting in a remarkable increase in sensor battery life from 13 to 997 days. However, it is essential to note that cost considerations play a pivotal role in this context. Embedded ML not only limits the advantages of collecting and storing copious raw data but also adds its own financial implications.

\section{Future Work Directions}
\label{sec:future}
The current state of smart healthcare has reached a level that was unimaginable just a few years ago, particularly due to the incorporation of IoT and ML into the process. However, there are still a number of challenges that need to be addressed, and there are also a great many areas that have not yet been discovered but that need to be put into practice as quickly as possible in order to improve this field even further. This section presents potential directions for the future based on the challenges that have been covered in the section that came before it.

\subsection{IoT based Systems}
Smart healthcare based on the IoT encounters various challenges, as extensively discussed in the previous section. This section outlines a comprehensive roadmap for future research based on the identified challenges.

\vspace{0.5 cm}
\noindent \textit{Using Predictive Maintenance Approach}\\
Implementing predictive maintenance procedures is an effective solution to tackle the challenges linked to maintenance costs. In \cite{liu2022probing}, authors have proposed a predictive maintenance process for IoT-enabled manufacturing. They use a convolutional neural network and long short-term memory for fault prediction and deep reinforcement learning to optimize production control and schedule maintenance. This approach aims to deliver more precise maintenance services, leading to an overall reduction in costs. Additionally, developing a dockerized blockchain client enables efficient resource utilization and optimized hardware configurations, contributing to lower maintenance costs and enhanced resource efficiency \cite{freire2022towards}.

\vspace{0.5 cm}
\noindent \textit{Developing DTN based System}\\
Singh et al. \cite{singh2022adaptive} have proposed the development of delay-tolerant networks (DTN) as a potential solution for network disruptions and latency issues in IoT systems. DTNs, designed to function in challenging conditions and across extensive distances, present a promising approach to effectively address both network stability and latency concerns. Furthermore, Namasudra et al. \cite{namasudra2023new} suggest that integrating ML models can enhance predictive capabilities, allowing for the anticipation of network faults before severe disruptions occur.

\vspace{0.5 cm}
\noindent \textit{Exploring Alternatives for Data Management}\\
Exploring alternative options for data handling is necessary due to the various challenges posed by existing methods. Hence, future research should focus on leveraging fog computing for data management, aiming for scalability and diverse data storage formats. Proposed solutions should include features like data replication to enhance availability for multiple user applications. It is crucial to address issues such as reliability, information validity, and overall performance, considering the limited resources of small IoT devices (e.g., storage, processing, and energy capacity) \cite{diene2020data}. Moreover, researchers have explored the integration of data trust methods to identify anomalies or untrustworthy data. This approach enhances data analysis and improves predictions in smart healthcare systems \cite{namasudra2023enhanced}.

\vspace{0.5 cm}
\noindent \textit{Introduce Optimization Techniques}\\
To address the issues of storage, energy, and computational limitations, significant efforts are required to optimize and develop effective energy protocols, particularly in fog computing systems, including virtual and ad hoc fog systems. These efforts should encompass aspects like network and computing resource optimization, as well as a shift toward environmentally friendly energies, such as renewable energy sources \cite{piovesan2018energy}. Moreover, implementing ResNet50 and MobileNetV2 as backbones with quantization techniques has proven to be an efficient and lightweight solution, contributing to the reduction of computational power in IoT-based systems \cite{10.1145/3526217}.

\vspace{0.5 cm}
\noindent \textit{Introduce Contact-less Approach for Data Collection}\\
To address the issue of identifying relevant data, a contactless approach, such as using cameras for data collection, can be introduced. The field of contactless methods holds substantial potential for future research, though it presents several unresolved challenges. Consequently, given the current scenario, a complete shift to a contactless method may not be a practical option. Hence, a more viable approach could involve a combination of both contact and contactless methods for enhanced reliability and success. Moreover, incorporating contactless feature extraction, along with signal processing techniques like PCA, ICA, filtering, and supervised ML methods, could offer a more robust solution \cite{malasinghe2019remote}.

\subsection{ML based Systems}
The challenges faced by the ML-based healthcare system have been explored in Section \ref{sec:ml_challenge}. This section provides some future research aspects based on these challenges to make the smart healthcare system more robust and efficient.

\vspace{0.6cm} 
\noindent \textit{Using DL method for Data Preparation}\\
Exploring future research opportunities, one potential area lies in utilizing solutions based on DL for addressing data preparation issues. One prominent example is in medical imaging, where the reliance on manual efforts for anomaly detection has been a longstanding practice. DL processes offer a promising solution by automating this detection process, thereby enhancing the accuracy of medical imaging \cite{kim2019deep}. However, the implementation of DL methods faces a major obstacle in the need for large volumes of data. Therefore, future research must address how DL can be seamlessly integrated while tackling existing issues or if alternative approaches may be more impactful in this context. Moreover, a crucial aspect is the inclusion of patient-specific factors in data aggregation. This can be achieved by integrating aggregators at the edge, facilitating a more personalized and context-aware healthcare approach. 

\vspace{0.5 cm}
\noindent \textit{Optimization techniques for High Dimentionality and Data Imbalance}\\
To mitigate challenges like high dimensionality in data and biases arising from data imbalance, upcoming research should concentrate on optimizing model architecture, enhancing training and validation procedures, and investigating sampling techniques or introducing ensemble learning \cite{yu2014improved}. It is crucial to ensure that the original data remains unaltered while addressing these challenges in order to enhance the effectiveness of ML in smart healthcare.

\vspace{0.5 cm}
\noindent \textit{Developing Scalable Platform for Big Data}\\
In order to tackle the challenges associated with handling extensive volumes of medical data, the development of scalable, robust, and elastic cloud platforms is imperative for effective big data management in the healthcare domain.
Moreover, future research endeavors could concentrate on implementing models based on MapReduce \cite{syed2019smart}. MapReduce-based models offer notable advantages, including higher scalability and improved performance through parallel processing. By leveraging the parallel processing capabilities inherent in MapReduce, healthcare applications can efficiently handle the massive datasets involved in medical analytics, paving the way for more effective and scalable ML solutions in the healthcare domain.

\vspace{0.5 cm}
\noindent \textit{Use Open and Interpretable AI}\\
Open and interpretable AI offers a potential solution to address legal, ethical, and social issues arising from ML in healthcare by focusing on enhancing the understandability of machine decisions \cite{vellido2020importance}. It aids in building trust, reducing bias, and enhancing human understanding. Open and interpretable AI addresses various challenges and issues, thereby improving the overall experience of service delivery, traceability, and confidence in the use of AI and ML tools in healthcare. Therefore, it is necessary to conduct research on interpretability and explanatory techniques for ML models within the Smart Healthcare Management (SHM) framework.

\subsection{IoT and ML based Systems}
In this section, we have discussed the future directions of IoT and ML-based smart healthcare challenges. 

\vspace{0.5cm}
\noindent\textit{Developing Lightweight and Energy Efficient Solution}\\
   Smart healthcare issues, such as high energy consumption and the lack of an effective solution for resource management, are of the utmost importance. As a result, it is essential to develop an innovative, lightweight, and energy-efficient ML-based data aggregation algorithm because the majority of the currently available solutions do not include these characteristics. In addition, it is necessary to develop innovative schemes that are able to divide the task among the various components of the IoT. These schemes should not only be able to solve the problem of resource constraints that are present in these networks, but they should also be able to provide a solution that has an acceptable level of accuracy \cite{khan2020challenges}. 

\vspace{0.5cm}

\noindent \textit{Continuous Learning for Handling Synchronization Issue}\\
Addressing model and data synchronization issues in smart healthcare requires future research efforts. Shah et al. \cite{shah2019artificial} have emphasized the need for ML algorithms to facilitate continuous learning from clinical data and apply this learning to new data. Implementing continuous learning is challenging in practice, and further research is needed to develop models that can emulate human brain-like thinking processes. Moreover, the exploration of fog computing-based solutions has gained popularity in recent times, presenting an avenue for further investigation to achieve improved results \cite{moghadas2020iot}.

\vspace{0.5cm}
\noindent\textit{Utilize Privacy-Enhancing Algorithms}\\
    Preserving the integrity and confidentiality of health records, which contain sensitive information and are susceptible to breaches, is of utmost importance \cite{namasudra2018security}. Therefore, it is necessary to develop energy-efficient and lightweight ML data aggregation algorithms that are also secure. It is necessary for the solutions to be able to preserve confidentiality and defend against breaches of privacy using a variety of data privacy protection algorithms \cite{das2022novel}. Integration of various data protection strategies, such as blockchain and differential privacy, is also open to consideration. This would result in improved data security. Developing improved access control methods is necessary to maintain the safety and security of the network. Additionally, the devices comprising the IoT should be designed to be tamper-resistant, safeguarding them from physical damage \cite{li2021comprehensive}.

\vspace{0.5 cm}
\noindent \textit{Designing Appropriate Devices}\\
Accurately distinguishing between true and false data collected via wearable devices presents a substantial hurdle in the healthcare domain. Additionally, adaptability is crucial for upcoming projects, especially in scenarios like movement detection where sensors must be versatile to accommodate the unique gait data of each individual. Exploring both contact-based and contactless methods, along with the development of DL models capable of discerning between genuine and false values, can be instrumental in addressing these challenges.

\vspace{0.5cm}
\noindent\textit{Ensuring Semantic Interoperability}\\
  Building a globally acknowledged standard and protocol infrastructure is an absolute necessity to tackle interoperability issues. One potential approach that could be taken in the future is to maintain the semantic interoperability of healthcare information \cite{li2021comprehensive}. This involves ensuring that when various healthcare systems or devices share information with one another, they are able to comprehend the meaning and context of the data in a manner that is both consistent and accurate. It is essential for effective communication and collaboration between various technologies used in the healthcare field that they speak the same language when exchanging important medical data.

\vspace{0.5cm}
\noindent\textit{Integrating ML and Optimization Methods}\\
    It is crucial to develop tools and methods for big data analytics that can perform analysis and extract the necessary information. Developing novel noise removal techniques is crucial for preprocessing the data, enabling more effective analysis and improving the data signal. More importantly, because IoT devices produce real-time data, utilizing ML techniques for real-time information monitoring and providing prompt responses is a very interesting future area for research. This could be a topic of study in the near future. In addition, the performance degradation of data aggregation techniques brought on by the topologies underlying them is a major cause for concern. It is necessary to conduct research on them in a dynamic and heterogeneous environment \cite{shokri2019review, xue2019using, khan2020secured}.

    \vspace{0.5 cm}
In addition to the aforementioned pathways, there are some broader areas for future research. For example, no comprehensive investigation has been conducted on these technologies to determine which big data technologies and ML techniques are most applicable to IoT healthcare. Furthermore, studies that interlink the two cross domains, i.e., big data analytics and healthcare, are still in their early stages, and as a result, the research community needs to pay even more attention to these types of studies.

\section{Conclusions}
\label{sec:conclusion}
In the 21\textsuperscript{st} century, IoT has thrived, enhancing daily decision-making and introducing services like pay-as-you-use models. The integration of IoT into modern life aims to improve quality of life by incorporating smart devices and technologies. Simultaneously, ML addresses business challenges through predictive analytics. In recent years, both ML and IoT have independently impacted healthcare, with ML focusing on algorithm development and IoT facilitating real-time monitoring. Researchers are now frequently exploring ML solutions within IoT-based healthcare, providing significant benefits in statistical and predictive analysis. This chapter provides a comprehensive introduction to IoT-based smart healthcare with ML, catering to newcomers interested in exploring this field. It covers research challenges and offers insights for future contributions. The chapter divides into three main sections: IoT-based smart healthcare, ML integration in traditional healthcare, and the synergy of both IoT and ML. It discusses real-life applications in these domains, addresses challenges in developing new architectural solutions, and outlines future research directions.

Overall, the chapter establishes a strong foundation for researchers keen on practical applications or innovative theoretical approaches in the realm of smart healthcare, providing a deep understanding of the challenges associated with ML and IoT applications. While the inherent and individual capabilities of IoT and ML create an ideal cross-domain synergy for developing IoT and ML-based smart healthcare, several challenges still exist. Firstly, there are  challenges faced by IoT and ML individually in healthcare and other domains. Secondly, issues arise with the cross-domain combination of IoT and ML, as both domains present various complexities. Therefore, future research should primarily address the inherent issues of each domain while also tackling the complexities of integrating these two domains. Considering the persistent research challenges, an appropriate approach could lead to the transformative changes in healthcare that the world anticipates.


\begingroup
\let\clearpage\relax 

\bibliographystyle{ieeetr}
\bibliography{main}

\begin{thebibliography}{100}

\bibitem{cook2012casas}
D.~J. Cook, A.~S. Crandall, B.~L. Thomas, and N.~C. Krishnan, ``Casas: A smart
  home in a box,'' {\em Computer}, vol.~46, no.~7, pp.~62--69, 2012.

\bibitem{alkhayyat2019wbsn}
A.~Alkhayyat, A.~A. Thabit, F.~A. Al-Mayali, Q.~H. Abbasi, {\em et~al.}, ``Wbsn
  in iot health-based application: toward delay and energy consumption
  minimization,'' {\em Journal of Sensors}, vol.~2019, 2019.

\bibitem{abdullah2015real}
A.~Abdullah, A.~Ismael, A.~Rashid, A.~Abou-ElNour, and M.~Tarique, ``Real time
  wireless health monitoring application using mobile devices,'' {\em
  International Journal of Computer Networks \& Communications (IJCNC)},
  vol.~7, no.~3, pp.~13--30, 2015.

\bibitem{wu2019wearable}
T.~Wu, J.-M. Redout{\'e}, and M.~Yuce, ``A wearable, low-power, real-time ecg
  monitor for smart t-shirt and iot healthcare applications,'' in {\em Advances
  in Body Area Networks I: Post-Conference Proceedings of BodyNets 2017},
  pp.~165--173, Springer, 2019.

\bibitem{fu2015system}
Y.~Fu and J.~Liu, ``System design for wearable blood oxygen saturation and
  pulse measurement device,'' {\em Procedia manufacturing}, vol.~3,
  pp.~1187--1194, 2015.

\bibitem{heshmat2018framework}
M.~Heshmat and A.-R.~S. Shehata, ``A framework about using internet of things
  for smart cancer treatment process,'' in {\em Proceedings of the
  international conference on industrial engineering and operations
  management}, pp.~1206--1211, 2018.

\bibitem{mahdavinejad2018machine}
M.~S. Mahdavinejad, M.~Rezvan, M.~Barekatain, P.~Adibi, P.~Barnaghi, and A.~P.
  Sheth, ``Machine learning for internet of things data analysis: A survey,''
  {\em Digital Communications and Networks}, vol.~4, no.~3, pp.~161--175, 2018.

\bibitem{qian2020orchestrating}
B.~Qian, J.~Su, Z.~Wen, D.~N. Jha, Y.~Li, Y.~Guan, D.~Puthal, P.~James,
  R.~Yang, A.~Y. Zomaya, {\em et~al.}, ``Orchestrating the development
  lifecycle of machine learning-based iot applications: A taxonomy and
  survey,'' {\em ACM Computing Surveys (CSUR)}, vol.~53, no.~4, pp.~1--47,
  2020.

\bibitem{babu2018survey}
G.~C. Babu and S.~Shantharajah, ``Survey on data analytics techniques in
  healthcare using iot platform,'' {\em International Journal of
  Reasoning-based Intelligent Systems}, vol.~10, no.~3-4, pp.~183--196, 2018.

\bibitem{ghazal2021iot}
T.~M. Ghazal, M.~K. Hasan, M.~T. Alshurideh, H.~M. Alzoubi, M.~Ahmad, S.~S.
  Akbar, B.~Al~Kurdi, and I.~A. Akour, ``Iot for smart cities: Machine learning
  approaches in smart healthcare—a review,'' {\em Future Internet}, vol.~13,
  no.~8, p.~218, 2021.

\bibitem{mohammadi2022applications}
F.~G. Mohammadi, F.~Shenavarmasouleh, and H.~R. Arabnia, ``Applications of
  machine learning in healthcare and internet of things (iot): a comprehensive
  review,'' {\em arXiv preprint arXiv:2202.02868}, 2022.

\bibitem{chawla2020ai}
N.~Chawla, ``Ai, iot and wearable technology for smart healthcare-a review.,''
  {\em International Journal of Recent Research Aspects}, vol.~7, no.~1, 2020.

\bibitem{alshehri2020comprehensive}
F.~Alshehri and G.~Muhammad, ``A comprehensive survey of the internet of things
  (iot) and ai-based smart healthcare,'' {\em IEEE Access}, vol.~9,
  pp.~3660--3678, 2020.

\bibitem{tunc2021survey}
M.~A. Tunc, E.~Gures, and I.~Shayea, ``A survey on iot smart healthcare:
  Emerging technologies, applications, challenges, and future trends,'' {\em
  arXiv preprint arXiv:2109.02042}, 2021.

\bibitem{yin2018smart}
H.~Yin, A.~O. Akmandor, A.~Mosenia, N.~K. Jha, {\em et~al.}, ``Smart
  healthcare,'' {\em Foundations and Trends{\textregistered} in Electronic
  Design Automation}, vol.~12, no.~4, pp.~401--466, 2018.

\bibitem{tian2019smart}
S.~Tian, W.~Yang, J.~M. Le~Grange, P.~Wang, W.~Huang, and Z.~Ye, ``Smart
  healthcare: making medical care more intelligent,'' {\em Global Health
  Journal}, vol.~3, no.~3, pp.~62--65, 2019.

\bibitem{liu2012internet}
M.-L. Liu, L.~Tao, and Z.~Yan, ``Internet of things-based electrocardiogram
  monitoring system,'' {\em Chinese Patent}, vol.~102, no.~764, p.~118, 2012.

\bibitem{agustine2018heart}
L.~Agustine, I.~Muljono, P.~R. Angka, A.~Gunadhi, D.~Lestariningsih, and W.~A.
  Weliamto, ``Heart rate monitoring device for arrhythmia using pulse oximeter
  sensor based on android,'' in {\em 2018 International Conference on Computer
  Engineering, Network and Intelligent Multimedia (CENIM)}, pp.~106--111, IEEE,
  2018.

\bibitem{cecil2018iomt}
J.~Cecil, A.~Gupta, M.~Pirela-Cruz, and P.~Ramanathan, ``An iomt based cyber
  training framework for orthopedic surgery using next generation internet
  technologies,'' {\em Informatics in Medicine Unlocked}, vol.~12,
  pp.~128--137, 2018.

\bibitem{su2020internet}
H.~Su, S.~E. Ovur, Z.~Li, Y.~Hu, J.~Li, A.~Knoll, G.~Ferrigno, and E.~De~Momi,
  ``Internet of things (iot)-based collaborative control of a redundant
  manipulator for teleoperated minimally invasive surgeries,'' in {\em 2020
  IEEE international conference on robotics and automation (ICRA)},
  pp.~9737--9742, IEEE, 2020.

\bibitem{yu2012posture}
M.~Yu, A.~Rhuma, S.~M. Naqvi, L.~Wang, and J.~Chambers, ``A posture
  recognition-based fall detection system for monitoring an elderly person in a
  smart home environment,'' {\em IEEE transactions on information technology in
  biomedicine}, vol.~16, no.~6, pp.~1274--1286, 2012.

\bibitem{estrin2010open}
D.~Estrin and I.~Sim, ``Open mhealth architecture: an engine for health care
  innovation,'' {\em Science}, vol.~330, no.~6005, pp.~759--760, 2010.

\bibitem{akmandor2017keep}
A.~O. Akmandor and N.~K. Jha, ``Keep the stress away with soda: Stress
  detection and alleviation system,'' {\em IEEE Transactions on Multi-Scale
  Computing Systems}, vol.~3, no.~4, pp.~269--282, 2017.

\bibitem{wahid2023covict}
M.~A. Wahid, S.~H.~R. Bukhari, A.~Daud, S.~E. Awan, and M.~A.~Z. Raja,
  ``Covict: an iot based architecture for covid-19 detection and contact
  tracing,'' {\em Journal of Ambient Intelligence and Humanized Computing},
  vol.~14, no.~6, pp.~7381--7398, 2023.

\bibitem{yin2017health}
H.~Yin and N.~K. Jha, ``A health decision support system for disease diagnosis
  based on wearable medical sensors and machine learning ensembles,'' {\em IEEE
  Transactions on Multi-Scale Computing Systems}, vol.~3, no.~4, pp.~228--241,
  2017.

\bibitem{merck2017chronic}
S.~F. Merck, ``Chronic disease and mobile technology: an innovative tool for
  clinicians,'' in {\em Nursing Forum}, vol.~52, pp.~298--305, Wiley Online
  Library, 2017.

\bibitem{willard2013effectiveness}
R.~Willard-Grace, D.~DeVore, E.~H. Chen, D.~Hessler, T.~Bodenheimer, and D.~H.
  Thom, ``The effectiveness of medical assistant health coaching for low-income
  patients with uncontrolled diabetes, hypertension, and hyperlipidemia:
  protocol for a randomized controlled trial and baseline characteristics of
  the study population,'' {\em BMC Family practice}, vol.~14, no.~1, pp.~1--10,
  2013.

\bibitem{andreu2015wearable}
J.~Andreu-Perez, D.~R. Leff, H.~M. Ip, and G.-Z. Yang, ``From wearable sensors
  to smart implants---toward pervasive and personalized healthcare,'' {\em IEEE
  Transactions on Biomedical Engineering}, vol.~62, no.~12, pp.~2750--2762,
  2015.

\bibitem{sundholm2014smart}
M.~Sundholm, J.~Cheng, B.~Zhou, A.~Sethi, and P.~Lukowicz, ``Smart-mat:
  Recognizing and counting gym exercises with low-cost resistive pressure
  sensing matrix,'' in {\em Proceedings of the 2014 ACM international joint
  conference on pervasive and ubiquitous computing}, pp.~373--382, 2014.

\bibitem{geller2016smart}
N.~L. Geller, D.-Y. Kim, and X.~Tian, ``Smart technology in lung disease
  clinical trials,'' {\em Chest}, vol.~149, no.~1, pp.~22--26, 2016.

\bibitem{hassan2018intelligent}
M.~K. Hassan, A.~I. El~Desouky, S.~M. Elghamrawy, and A.~M. Sarhan,
  ``Intelligent hybrid remote patient-monitoring model with cloud-based
  framework for knowledge discovery,'' {\em Computers \& Electrical
  Engineering}, vol.~70, pp.~1034--1048, 2018.

\bibitem{hassan2019hybrid}
M.~K. Hassan, A.~I. El~Desouky, S.~M. Elghamrawy, and A.~M. Sarhan, ``A hybrid
  real-time remote monitoring framework with nb-woa algorithm for patients with
  chronic diseases,'' {\em Future Generation Computer Systems}, vol.~93,
  pp.~77--95, 2019.

\bibitem{syed2019smart}
L.~Syed, S.~Jabeen, S.~Manimala, and A.~Alsaeedi, ``Smart healthcare framework
  for ambient assisted living using iomt and big data analytics techniques,''
  {\em Future Generation Computer Systems}, vol.~101, pp.~136--151, 2019.

\bibitem{chatrati2022smart}
S.~P. Chatrati, G.~Hossain, A.~Goyal, A.~Bhan, S.~Bhattacharya, D.~Gaurav, and
  S.~M. Tiwari, ``Smart home health monitoring system for predicting type 2
  diabetes and hypertension,'' {\em Journal of King Saud University-Computer
  and Information Sciences}, vol.~34, no.~3, pp.~862--870, 2022.

\bibitem{pham2017predicting}
T.~Pham, T.~Tran, D.~Phung, and S.~Venkatesh, ``Predicting healthcare
  trajectories from medical records: A deep learning approach,'' {\em Journal
  of biomedical informatics}, vol.~69, pp.~218--229, 2017.

\bibitem{vijayakumar2019fog}
V.~Vijayakumar, D.~Malathi, V.~Subramaniyaswamy, P.~Saravanan, and R.~Logesh,
  ``Fog computing-based intelligent healthcare system for the detection and
  prevention of mosquito-borne diseases,'' {\em Computers in Human Behavior},
  vol.~100, pp.~275--285, 2019.

\bibitem{precisionmedicine}
[Online; accessed November 22, 2023].

\bibitem{dong2015anticancer}
Z.~Dong, N.~Zhang, C.~Li, H.~Wang, Y.~Fang, J.~Wang, and X.~Zheng, ``Anticancer
  drug sensitivity prediction in cell lines from baseline gene expression
  through recursive feature selection,'' {\em BMC cancer}, vol.~15, no.~1,
  pp.~1--12, 2015.

\bibitem{kalinin2018deep}
A.~A. Kalinin, G.~A. Higgins, N.~Reamaroon, S.~Soroushmehr, A.~Allyn-Feuer,
  I.~D. Dinov, K.~Najarian, and B.~D. Athey, ``Deep learning in
  pharmacogenomics: from gene regulation to patient stratification,'' {\em
  Pharmacogenomics}, vol.~19, no.~7, pp.~629--650, 2018.

\bibitem{das2023aespnet}
S.~K. Das, S.~Namasudra, A.~Kumar, and N.~R. Moparthi, ``Aespnet: Attention
  enhanced stacked parallel network to improve automatic diabetic foot ulcer
  identification,'' {\em Image and Vision Computing}, vol.~138, p.~104809,
  2023.

\bibitem{bhatia2017comprehensive}
M.~Bhatia and S.~K. Sood, ``A comprehensive health assessment framework to
  facilitate iot-assisted smart workouts: A predictive healthcare
  perspective,'' {\em Computers in Industry}, vol.~92, pp.~50--66, 2017.

\bibitem{joshi2021healthcare}
S.~Joshi, H.~Kumar, J.~Babu, A.~Raju, and M.~Nihaz, ``Healthcare assistant—a
  tool to predict disease using machine learning,'' in {\em International
  Conference on Micro-Electronics and Telecommunication Engineering},
  pp.~221--229, Springer, 2021.

\bibitem{farahani2020healthcare}
B.~Farahani, F.~Firouzi, and K.~Chakrabarty, ``Healthcare iot,'' {\em
  Intelligent Internet of Things: From Device to Fog and Cloud}, pp.~515--545,
  2020.

\bibitem{tuli2020healthfog}
S.~Tuli, N.~Basumatary, S.~S. Gill, M.~Kahani, R.~C. Arya, G.~S. Wander, and
  R.~Buyya, ``Healthfog: An ensemble deep learning based smart healthcare
  system for automatic diagnosis of heart diseases in integrated iot and fog
  computing environments,'' {\em Future Generation Computer Systems}, vol.~104,
  pp.~187--200, 2020.

\bibitem{ahmad2020review}
T.~Ahmad and H.~Chen, ``A review on machine learning forecasting growth trends
  and their real-time applications in different energy systems,'' {\em
  Sustainable Cities and Society}, vol.~54, p.~102010, 2020.

\bibitem{asthana2017recommendation}
S.~Asthana, A.~Megahed, and R.~Strong, ``A recommendation system for proactive
  health monitoring using iot and wearable technologies,'' in {\em 2017 IEEE
  international conference on AI \& mobile services (AIMS)}, pp.~14--21, IEEE,
  2017.

\bibitem{subramaniyaswamy2019ontology}
V.~Subramaniyaswamy, G.~Manogaran, R.~Logesh, V.~Vijayakumar, N.~Chilamkurti,
  D.~Malathi, and N.~Senthilselvan, ``An ontology-driven personalized food
  recommendation in iot-based healthcare system,'' {\em The Journal of
  Supercomputing}, vol.~75, pp.~3184--3216, 2019.

\bibitem{qiu2016efficient}
T.~Qiu, X.~Liu, L.~Feng, Y.~Zhou, and K.~Zheng, ``An efficient tree-based
  self-organizing protocol for internet of things,'' {\em Ieee Access}, vol.~4,
  pp.~3535--3546, 2016.

\bibitem{alsheikh2016rate}
M.~A. Alsheikh, S.~Lin, D.~Niyato, and H.-P. Tan, ``Rate-distortion balanced
  data compression for wireless sensor networks,'' {\em IEEE Sensors Journal},
  vol.~16, no.~12, pp.~5072--5083, 2016.

\bibitem{pan2020enhanced}
Y.~Pan, M.~Fu, B.~Cheng, X.~Tao, and J.~Guo, ``Enhanced deep learning assisted
  convolutional neural network for heart disease prediction on the internet of
  medical things platform,'' {\em Ieee Access}, vol.~8, pp.~189503--189512,
  2020.

\bibitem{pradhan2020medical}
K.~Pradhan and P.~Chawla, ``Medical internet of things using machine learning
  algorithms for lung cancer detection,'' {\em Journal of Management
  Analytics}, vol.~7, no.~4, pp.~591--623, 2020.

\bibitem{negra2018wban}
R.~Negra, I.~Jemili, A.~Zemmari, M.~Mosbah, and A.~Belghith, ``Wban path loss
  based approach for human activity recognition with machine learning
  techniques,'' in {\em 2018 14th International Wireless Communications \&
  Mobile Computing Conference (IWCMC)}, pp.~470--475, IEEE, 2018.

\bibitem{matar2016internet}
G.~Matar, J.-M. Lina, J.~Carrier, A.~Riley, and G.~Kaddoum, ``Internet of
  things in sleep monitoring: An application for posture recognition using
  supervised learning,'' in {\em 2016 IEEE 18th International conference on
  e-Health networking, applications and services (Healthcom)}, pp.~1--6, IEEE,
  2016.

\bibitem{gulati2022friendcare}
N.~Gulati and P.~D. Kaur, ``Friendcare-aal: A robust social iot based alert
  generation system for ambient assisted living,'' {\em Journal of Ambient
  Intelligence and Humanized Computing}, pp.~1--28, 2022.

\bibitem{rupasinghe2022towards}
I.~Rupasinghe and M.~Maduranga, ``Towards ambient assisted living (aal): Design
  of an iotbased elderly activity monitoring system,'' {\em International
  Journal of Engineering and Manufacturing (IJEM)}, vol.~12, no.~2, pp.~1--10,
  2022.

\bibitem{khan2018performance}
F.~Khan, A.~ur~Rehman, M.~Usman, Z.~Tan, and D.~Puthal, ``Performance of
  cognitive radio sensor networks using hybrid automatic repeat request:
  Stop-and-wait,'' {\em Mobile Networks and Applications}, vol.~23,
  pp.~479--488, 2018.

\bibitem{gope2015bsn}
P.~Gope and T.~Hwang, ``Bsn-care: A secure iot-based modern healthcare system
  using body sensor network,'' {\em IEEE sensors journal}, vol.~16, no.~5,
  pp.~1368--1376, 2015.

\bibitem{elsaadany2017wireless}
Y.~ElSaadany, A.~J.~A. Majumder, and D.~R. Ucci, ``A wireless early prediction
  system of cardiac arrest through iot,'' in {\em 2017 IEEE 41st annual
  computer software and applications conference (COMPSAC)}, vol.~2,
  pp.~690--695, IEEE, 2017.

\bibitem{wood2015toward}
T.~Wood, K.~Ramakrishnan, J.~Hwang, G.~Liu, and W.~Zhang, ``Toward a
  software-based network: integrating software defined networking and network
  function virtualization,'' {\em IEEE Network}, vol.~29, no.~3, pp.~36--41,
  2015.

\bibitem{satija2017real}
U.~Satija, B.~Ramkumar, and M.~S. Manikandan, ``Real-time signal quality-aware
  ecg telemetry system for iot-based health care monitoring,'' {\em IEEE
  Internet of Things Journal}, vol.~4, no.~3, pp.~815--823, 2017.

\bibitem{jan2019smartedge}
M.~A. Jan, W.~Zhang, M.~Usman, Z.~Tan, F.~Khan, and E.~Luo, ``Smartedge: An
  end-to-end encryption framework for an edge-enabled smart city application,''
  {\em Journal of Network and Computer Applications}, vol.~137, pp.~1--10,
  2019.

\bibitem{kanagasabai2016brain}
P.~S. Kanagasabai, R.~Gautam, and G.~Rathna, ``Brain-computer interface
  learning system for quadriplegics,'' in {\em 2016 IEEE 4th international
  conference on MOOCs, innovation and technology in education (MITE)},
  pp.~258--262, IEEE, 2016.

\bibitem{tissaoui2020uncertainty}
A.~Tissaoui and M.~Saidi, ``Uncertainty in iot for smart healthcare:
  Challenges, and opportunities,'' in {\em The Impact of Digital Technologies
  on Public Health in Developed and Developing Countries: 18th International
  Conference, ICOST 2020, Hammamet, Tunisia, June 24--26, 2020, Proceedings
  18}, pp.~232--239, Springer, 2020.

\bibitem{malasinghe2019remote}
L.~P. Malasinghe, N.~Ramzan, and K.~Dahal, ``Remote patient monitoring: a
  comprehensive study,'' {\em Journal of Ambient Intelligence and Humanized
  Computing}, vol.~10, pp.~57--76, 2019.

\bibitem{hu2015efficient}
Y.-H. Hu, W.-C. Lin, C.-F. Tsai, S.-W. Ke, and C.-W. Chen, ``An efficient data
  preprocessing approach for large scale medical data mining,'' {\em Technology
  and Health Care}, vol.~23, no.~2, pp.~153--160, 2015.

\bibitem{bellazzi2008predictive}
R.~Bellazzi and B.~Zupan, ``Predictive data mining in clinical medicine:
  current issues and guidelines,'' {\em International journal of medical
  informatics}, vol.~77, no.~2, pp.~81--97, 2008.

\bibitem{chen2019develop}
P.-H.~C. Chen, Y.~Liu, and L.~Peng, ``How to develop machine learning models
  for healthcare,'' {\em Nature materials}, vol.~18, no.~5, pp.~410--414, 2019.

\bibitem{amador2022early}
T.~Amador, S.~Saturnino, A.~Veloso, and N.~Ziviani, ``Early identification of
  icu patients at risk of complications: Regularization based on robustness and
  stability of explanations,'' {\em Artificial Intelligence in Medicine},
  vol.~128, p.~102283, 2022.

\bibitem{naha2020deadline}
R.~K. Naha, S.~Garg, A.~Chan, and S.~K. Battula, ``Deadline-based dynamic
  resource allocation and provisioning algorithms in fog-cloud environment,''
  {\em Future Generation Computer Systems}, vol.~104, pp.~131--141, 2020.

\bibitem{zhou2017security}
J.~Zhou, Z.~Cao, X.~Dong, and A.~V. Vasilakos, ``Security and privacy for
  cloud-based iot: Challenges,'' {\em IEEE Communications Magazine}, vol.~55,
  no.~1, pp.~26--33, 2017.

\bibitem{ali2020resource}
S.~A. Ali, M.~Ansari, and M.~Alam, ``Resource management techniques for
  cloud-based iot environment,'' {\em Internet of Things (IoT) Concepts and
  Applications}, pp.~63--87, 2020.

\bibitem{khan2020challenges}
I.~H. Khan, M.~I. Khan, and S.~Khan, ``Challenges of iot implementation in
  smart city development,'' in {\em Smart Cities—Opportunities and
  Challenges: Select Proceedings of ICSC 2019}, pp.~475--486, Springer, 2020.

\bibitem{sharma2020performance}
D.~Sharma and R.~Tripathi, ``Performance of internet of things based healthcare
  secure services and its importance: Issue and challenges,'' tech. rep.,
  Technical report, EasyChair, 2020.

\bibitem{jan2019payload}
M.~A. Jan, F.~Khan, M.~Alam, and M.~Usman, ``A payload-based mutual
  authentication scheme for internet of things,'' {\em Future Generation
  Computer Systems}, vol.~92, pp.~1028--1039, 2019.

\bibitem{flynn2020knock}
T.~Flynn, G.~Grispos, W.~Glisson, and W.~Mahoney, ``Knock! knock! who is there?
  investigating data leakage from a medical internet of things hijacking
  attack,'' 2020.

\bibitem{williams2016always}
P.~A. Williams and V.~McCauley, ``Always connected: The security challenges of
  the healthcare internet of things,'' in {\em 2016 IEEE 3rd World Forum on
  Internet of Things (WF-IoT)}, pp.~30--35, IEEE, 2016.

\bibitem{khan2014fairness}
F.~Khan, ``Fairness and throughput improvement in multihop wireless ad hoc
  networks,'' in {\em 2014 IEEE 27th Canadian Conference on Electrical and
  Computer Engineering (CCECE)}, pp.~1--6, IEEE, 2014.

\bibitem{qadri2020future}
Y.~A. Qadri, A.~Nauman, Y.~B. Zikria, A.~V. Vasilakos, and S.~W. Kim, ``The
  future of healthcare internet of things: a survey of emerging technologies,''
  {\em IEEE Communications Surveys \& Tutorials}, vol.~22, no.~2,
  pp.~1121--1167, 2020.

\bibitem{park2020energy}
J.~Park, G.~Bhat, A.~Nk, C.~S. Geyik, U.~Y. Ogras, and H.~G. Lee, ``Energy per
  operation optimization for energy-harvesting wearable iot devices,'' {\em
  Sensors}, vol.~20, no.~3, p.~764, 2020.

\bibitem{abbasian2020survey}
S.~Abbasian~Dehkordi, K.~Farajzadeh, J.~Rezazadeh, R.~Farahbakhsh,
  K.~Sandrasegaran, and M.~Abbasian~Dehkordi, ``A survey on data aggregation
  techniques in iot sensor networks,'' {\em Wireless Networks}, vol.~26,
  pp.~1243--1263, 2020.

\bibitem{selvaraj2020challenges}
S.~Selvaraj and S.~Sundaravaradhan, ``Challenges and opportunities in iot
  healthcare systems: a systematic review,'' {\em SN Applied Sciences}, vol.~2,
  no.~1, p.~139, 2020.

\bibitem{mittal2019energy}
M.~Mittal, S.~Tanwar, B.~Agarwal, and L.~M. Goyal, ``Energy conservation for
  iot devices,'' {\em Concepts, Paradigms and Solutions, Studies in Systems,
  Decision and Control, in Preparation}, pp.~1--365, 2019.

\bibitem{gill2019bio}
S.~S. Gill and R.~Buyya, ``Bio-inspired algorithms for big data analytics: a
  survey, taxonomy, and open challenges,'' in {\em Big data analytics for
  intelligent healthcare management}, pp.~1--17, Elsevier, 2019.

\bibitem{wan2019similarity}
R.~Wan, N.~Xiong, Q.~Hu, H.~Wang, and J.~Shang, ``Similarity-aware data
  aggregation using fuzzy c-means approach for wireless sensor networks,'' {\em
  EURASIP Journal on Wireless Communications and Networking}, vol.~2019,
  pp.~1--11, 2019.

\bibitem{qi2019convolutional}
G.~Qi, H.~Wang, M.~Haner, C.~Weng, S.~Chen, and Z.~Zhu, ``Convolutional neural
  network based detection and judgement of environmental obstacle in vehicle
  operation,'' {\em CAAI Transactions on Intelligence Technology}, vol.~4,
  no.~2, pp.~80--91, 2019.

\bibitem{li2020secrecy}
X.~Li, M.~Zhao, Y.~Liu, L.~Li, Z.~Ding, and A.~Nallanathan, ``Secrecy analysis
  of ambient backscatter noma systems under i/q imbalance,'' {\em IEEE
  Transactions on Vehicular Technology}, vol.~69, no.~10, pp.~12286--12290,
  2020.

\bibitem{wiens2019engine}
T.~Wiens, ``Engine speed reduction for hydraulic machinery using predictive
  algorithms,'' {\em International Journal of Hydromechatronics}, vol.~2,
  no.~1, pp.~16--31, 2019.

\bibitem{li2020uav}
X.~Li, Q.~Wang, Y.~Liu, T.~A. Tsiftsis, Z.~Ding, and A.~Nallanathan,
  ``Uav-aided multi-way noma networks with residual hardware impairments,''
  {\em IEEE Wireless Communications Letters}, vol.~9, no.~9, pp.~1538--1542,
  2020.

\bibitem{shokri2019review}
M.~Shokri and K.~Tavakoli, ``A review on the artificial neural network approach
  to analysis and prediction of seismic damage in infrastructure,'' {\em
  International Journal of Hydromechatronics}, vol.~2, no.~4, pp.~178--196,
  2019.

\bibitem{xue2019using}
X.~Xue, J.~Lu, and J.~Chen, ``Using nsga-iii for optimising biomedical ontology
  alignment,'' {\em CAAI Transactions on Intelligence Technology}, vol.~4,
  no.~3, pp.~135--141, 2019.

\bibitem{ma2019numerical}
J.~Ma, ``Numerical modelling of underwater structural impact damage problems
  based on the material point method,'' {\em International Journal of
  Hydromechatronics}, vol.~2, no.~4, pp.~99--110, 2019.

\bibitem{khan2020secured}
F.~Khan, A.~ur~Rehman, and M.~A. Jan, ``A secured and reliable communication
  scheme in cognitive hybrid arq-aided smart city,'' {\em Computers \&
  Electrical Engineering}, vol.~81, p.~106502, 2020.

\bibitem{tingting2019three}
Y.~Tingting, W.~Junqian, W.~Lintai, and X.~Yong, ``Three-stage network for age
  estimation,'' {\em CAAI Transactions on Intelligence Technology}, vol.~4,
  no.~2, pp.~122--126, 2019.

\bibitem{ishtiaq2019performance}
M.~Ishtiaq, A.~U. Rehman, F.~Khan, A.~Salam, {\em et~al.}, ``Performance
  investigation of sr-harq transmission scheme in realistic cognitive radio
  system,'' in {\em 2019 IEEE 9th Annual Computing and Communication Workshop
  and Conference (CCWC)}, pp.~0258--0263, IEEE, 2019.

\bibitem{hussain2020machine}
F.~Hussain, S.~A. Hassan, R.~Hussain, and E.~Hossain, ``Machine learning for
  resource management in cellular and iot networks: Potentials, current
  solutions, and open challenges,'' {\em IEEE communications surveys \&
  tutorials}, vol.~22, no.~2, pp.~1251--1275, 2020.

\bibitem{hassabis2017neuroscience}
D.~Hassabis, D.~Kumaran, C.~Summerfield, and M.~Botvinick,
  ``Neuroscience-inspired artificial intelligence,'' {\em Neuron}, vol.~95,
  no.~2, pp.~245--258, 2017.

\bibitem{kaur2020internet}
H.~Kaur, M.~Atif, and R.~Chauhan, ``An internet of healthcare things
  (ioht)-based healthcare monitoring system,'' in {\em Advances in Intelligent
  Computing and Communication: Proceedings of ICAC 2019}, pp.~475--482,
  Springer, 2020.

\bibitem{almolhis2020security}
N.~Almolhis, A.~M. Alashjaee, S.~Duraibi, F.~Alqahtani, and A.~N. Moussa, ``The
  security issues in iot-cloud: a review,'' in {\em 2020 16th IEEE
  International Colloquium on Signal Processing \& Its Applications (CSPA)},
  pp.~191--196, IEEE, 2020.

\bibitem{bansal2020iot}
S.~Bansal and D.~Kumar, ``Iot ecosystem: A survey on devices, gateways,
  operating systems, middleware and communication,'' {\em International Journal
  of Wireless Information Networks}, vol.~27, pp.~340--364, 2020.

\bibitem{bhattacharjya2020present}
A.~Bhattacharjya, X.~Zhong, J.~Wang, and X.~Li, ``Present scenarios of iot
  projects with security aspects focused,'' {\em Digital Twin Technologies and
  Smart Cities}, pp.~95--122, 2020.

\bibitem{yang2020federated}
K.~Yang, Y.~Shi, Y.~Zhou, Z.~Yang, L.~Fu, and W.~Chen, ``Federated machine
  learning for intelligent iot via reconfigurable intelligent surface,'' {\em
  IEEE network}, vol.~34, no.~5, pp.~16--22, 2020.

\bibitem{hsu2017fallcare+}
C.~C.-H. Hsu, M.~Y.-C. Wang, H.~C. Shen, R.~H.-C. Chiang, and C.~H. Wen,
  ``Fallcare+: An iot surveillance system for fall detection,'' in {\em 2017
  International conference on applied system innovation (ICASI)}, pp.~921--922,
  IEEE, 2017.

\bibitem{zhang2018internet}
X.~Zhang, L.~Yao, S.~Zhang, S.~Kanhere, M.~Sheng, and Y.~Liu, ``Internet of
  things meets brain--computer interface: A unified deep learning framework for
  enabling human-thing cognitive interactivity,'' {\em IEEE Internet of Things
  Journal}, vol.~6, no.~2, pp.~2084--2092, 2018.

\bibitem{de2016wearable}
D.~de~Arruda and G.~P. Hancke, ``Wearable device localisation using machine
  learning techniques,'' in {\em 2016 IEEE 25th International symposium on
  industrial electronics (ISIE)}, pp.~1110--1115, IEEE, 2016.

\bibitem{ara2017case}
A.~Ara and A.~Ara, ``Case study: Integrating iot, streaming analytics and
  machine learning to improve intelligent diabetes management system,'' in {\em
  2017 International conference on energy, communication, data analytics and
  soft computing (ICECDS)}, pp.~3179--3182, IEEE, 2017.

\bibitem{fafoutis2018extending}
X.~Fafoutis, L.~Marchegiani, A.~Elsts, J.~Pope, R.~Piechocki, and I.~Craddock,
  ``Extending the battery lifetime of wearable sensors with embedded machine
  learning,'' in {\em 2018 IEEE 4th World Forum on Internet of Things
  (WF-IoT)}, pp.~269--274, IEEE, 2018.

\bibitem{liu2022probing}
C.~Liu, H.~Zhu, D.~Tang, Q.~Nie, T.~Zhou, L.~Wang, and Y.~Song, ``Probing an
  intelligent predictive maintenance approach with deep learning and augmented
  reality for machine tools in iot-enabled manufacturing,'' {\em Robotics and
  Computer-Integrated Manufacturing}, vol.~77, p.~102357, 2022.

\bibitem{freire2022towards}
W.~P. Freire, W.~S. Melo~Jr, V.~D. do~Nascimento, P.~R. Nascimento, and A.~O.
  de~S{\'a}, ``Towards a secure and scalable maritime monitoring system using
  blockchain and low-cost iot technology,'' {\em Sensors}, vol.~22, no.~13,
  p.~4895, 2022.

\bibitem{singh2022adaptive}
A.~K. Singh, R.~Pamula, and G.~Srivastava, ``An adaptive energy aware dtn-based
  communication layer for cyber-physical systems,'' {\em Sustainable Computing:
  Informatics and Systems}, vol.~35, p.~100657, 2022.

\bibitem{namasudra2023new}
S.~Namasudra, P.~Lorenz, and U.~Ghosh, ``The new era of computer network by
  using machine learning,'' {\em Mobile Networks and Applications}, pp.~1--3,
  2023.

\bibitem{diene2020data}
B.~Di{\`e}ne, J.~J. Rodrigues, O.~Diallo, E.~H.~M. Ndoye, and V.~V. Korotaev,
  ``Data management techniques for internet of things,'' {\em Mechanical
  Systems and Signal Processing}, vol.~138, p.~106564, 2020.

\bibitem{namasudra2023enhanced}
S.~Namasudra, S.~Dhamodharavadhani, R.~Rathipriya, R.~G. Crespo, and N.~R.
  Moparthi, ``Enhanced neural network-based univariate time-series forecasting
  model for big data,'' {\em Big Data}, 2023.

\bibitem{piovesan2018energy}
N.~Piovesan, A.~F. Gambin, M.~Miozzo, M.~Rossi, and P.~Dini, ``Energy
  sustainable paradigms and methods for future mobile networks: A survey,''
  {\em Computer Communications}, vol.~119, pp.~101--117, 2018.

\bibitem{10.1145/3526217}
K.~Manjari, M.~Verma, G.~Singal, and S.~Namasudra, ``Qest: Quantized and
  efficient scene text detector using deep learning,'' {\em ACM Trans. Asian
  Low-Resour. Lang. Inf. Process.}, vol.~22, may 2023.

\bibitem{kim2019deep}
M.~Kim, J.~Yun, Y.~Cho, K.~Shin, R.~Jang, H.-j. Bae, and N.~Kim, ``Deep
  learning in medical imaging,'' {\em Neurospine}, vol.~16, no.~4, p.~657,
  2019.

\bibitem{amann2020explainability}
J.~Amann, A.~Blasimme, E.~Vayena, D.~Frey, V.~I. Madai, and P.~Consortium,
  ``Explainability for artificial intelligence in healthcare: a
  multidisciplinary perspective,'' {\em BMC medical informatics and decision
  making}, vol.~20, pp.~1--9, 2020.

\bibitem{vellido2020importance}
A.~Vellido, ``The importance of interpretability and visualization in machine
  learning for applications in medicine and health care,'' {\em Neural
  computing and applications}, vol.~32, no.~24, pp.~18069--18083, 2020.

\bibitem{yu2014improved}
H.~Yu and J.~Ni, ``An improved ensemble learning method for classifying
  high-dimensional and imbalanced biomedicine data,'' {\em IEEE/ACM
  transactions on computational biology and bioinformatics}, vol.~11, no.~4,
  pp.~657--666, 2014.

\bibitem{namasudra2018security}
S.~Namasudra, D.~Devi, S.~Choudhary, R.~Patan, and S.~Kallam, ``Security,
  privacy, trust, and anonymity,'' {\em Advances of DNA computing in
  cryptography}, vol.~1, pp.~138--150, 2018.

\bibitem{das2022novel}
S.~Das and S.~Namasudra, ``A novel hybrid encryption method to secure
  healthcare data in iot-enabled healthcare infrastructure,'' {\em Computers
  and Electrical Engineering}, vol.~101, p.~107991, 2022.

\bibitem{li2021comprehensive}
W.~Li, Y.~Chai, F.~Khan, S.~R.~U. Jan, S.~Verma, V.~G. Menon, f.~Kavita, and
  X.~Li, ``A comprehensive survey on machine learning-based big data analytics
  for iot-enabled smart healthcare system,'' {\em Mobile networks and
  applications}, vol.~26, pp.~234--252, 2021.

\bibitem{shah2019artificial}
P.~Shah, F.~Kendall, S.~Khozin, R.~Goosen, J.~Hu, J.~Laramie, M.~Ringel, and
  N.~Schork, ``Artificial intelligence and machine learning in clinical
  development: a translational perspective,'' {\em NPJ digital medicine},
  vol.~2, no.~1, p.~69, 2019.

\bibitem{moghadas2020iot}
E.~Moghadas, J.~Rezazadeh, and R.~Farahbakhsh, ``An iot patient monitoring
  based on fog computing and data mining: Cardiac arrhythmia usecase,'' {\em
  Internet of Things}, vol.~11, p.~100251, 2020.

\end{thebibliography}

\endgroup

\end{document}